\documentclass[prd,twocolumn,showpacs,floatfix,amsmath,nofootinbib,amssymb,floatfix]{revtex4}
\usepackage{graphicx,color,dcolumn,booktabs,bm}
\usepackage{longtable,lscape}
\usepackage{txfonts}
\usepackage{overpic}
\usepackage{amssymb}
\usepackage{indentfirst}
\usepackage{feynmf}   %{feynmp}
\usepackage{slashed}  %for Feynman symbols
\usepackage{cases}
\usepackage{color}
\usepackage{multirow}
\usepackage{epstopdf}
\usepackage{graphicx,color,dcolumn,booktabs,bm}
\usepackage[colorlinks, citecolor=blue,anchorcolor=red,menucolor=red, linkcolor=red,filecolor=red,runcolor=red,urlcolor=blue,frenchlinks=red]{hyperref}

\graphicspath{{Figures/}} %

\begin{document}
%\begin{CJK}{GBK}{}

\title{Higher radial and orbital excitations in the charmed meson family}

\author{Qin-Tao Song$^{1,2,4}$}\email{songqint@impcas.ac.cn}
\author{Dian-Yong Chen$^{1,2}$\footnote{Corresponding author}}\email{chendy@impcas.ac.cn}
\author{Xiang Liu$^{2,3}$\footnote{Corresponding author}}\email{xiangliu@lzu.edu.cn}
\author{Takayuki Matsuki$^{5,6}$}\email{matsuki@tokyo-kasei.ac.jp}
\affiliation{$^1$Nuclear Theory Group, Institute of Modern Physics of CAS,
Lanzhou 730000, China\\
$^2$Research Center for Hadron
and CSR Physics, Lanzhou University $\&$ Institute of Modern Physics
of CAS,
Lanzhou 730000, China\\
$^3$School of Physical Science and Technology, Lanzhou University,
Lanzhou 730000, China\\
$^4$University of Chinese Academy of Sciences, Beijing 100049, China\\
$^5$Tokyo Kasei University, 1-18-1 Kaga, Itabashi, Tokyo 173-8602, Japan\\
$^6$Theoretical Research Division, Nishina Center, RIKEN, Saitama 351-0198, Japan}

\begin{abstract}

Having abundant experimental information of charmed mesons together with the present research status, we systematically study higher radial and orbital excitations in the charmed meson family by analyzing the mass spectrum and by calculating their OZI-allowed two-body decay behaviors. This phenomenological analysis
% of the higher radial and orbital excitations in the charmed meson family
reveals underlying properties of the newly observed charmed states $D(2550)$, $D^*(2600)$, $D^*(2760)$, $D(2750)$, $D_J(2580)$, $D^*_J(2650)$, $D^*_J(2760)$, $D_J(2740)$, $D_J(3000)$ and $D^*_J(3000)$ to provide valuable information of the charmed mesons still missing in experiments.

\end{abstract}

 \maketitle

\section{introduction}\label{sec1}

\renewcommand{\arraystretch}{1.3}
\begin{table*}[htbp]
\caption{Experimental information of the observed charmed mesons. \label{table:review} }
\centering
\begin{tabular}{cccccc}
\toprule[1pt] State & Mass (MeV) \cite{Beringer:1900zz} & Width (MeV) \cite{Beringer:1900zz} &
$1^{st}$ observation &Observed decay modes\\ \midrule[1pt]
$D$ & $1864.84\pm0.07$ &   & &\\

$D^{\ast}$ & $2010.26\pm0.07$& $0.083$ & &\\

$D_{1}(2420)$ & $2421.4\pm0.6$ & $70\pm21$ & ARGUS \cite{Albrecht:1985as} &$D^* \pi$ \\
$D_{2}^*(2460)$ & $2464.3\pm1.6$ & $20\pm10\pm5$ & TPS   \cite{Anjos:1988uf}&$D^+\pi^-$  \\

$D_{1}(2430)$ & $2427\pm26\pm25$ & $384^{+107}_{-75}\pm75$ & Belle \cite{Abe:2003zm} &$D^* \pi$ \\
& $2477\pm28$& $266\pm97$ & BaBar \cite{Aubert:2006zb} &$D^* \pi$ \\

$D_{0}^*(2400)$ & $2308\pm17\pm32$ & $276\pm21\pm63$ &Belle \cite{Abe:2003zm}&  $D \pi$ \\

& $2407\pm21\pm35$ & $240\pm55\pm59$ &FOCUS \cite{Link:2003bd}&  $D \pi$ \\
& $2297\pm8\pm20$ & $273\pm12\pm48$ &BaBar \cite{Aubert:2009wg} & $D \pi$ \\

$D(2550)$ & $2539.4\pm4.5\pm6.8$  & $130\pm12\pm13$ &BaBar \cite{delAmoSanchez:2010vq}&  $D^* \pi$ \\
$D^*(2600)$ & $2608.7\pm2.4\pm2.5$ & $93\pm6\pm13$ &BaBar \cite{delAmoSanchez:2010vq}&  $D^{(*)} \pi$ \\

$D(2750)$ & $2752.4\pm1.7\pm2.7$ & $71\pm6\pm11$ &BaBar \cite{delAmoSanchez:2010vq}&  $D^{*} \pi$ \\

$D^*(2760)$ & $2763.3\pm2.3\pm2.3$ & $60.9\pm5.1\pm3.6$ &BaBar \cite{delAmoSanchez:2010vq}&  $D \pi$ \\

$D(2580)$ & $2579.5\pm3.4\pm5.5$ & $177.5\pm17.8\pm46.0$ &LHCb \cite{Aaij:2013sza}&  $D^* \pi$ \\
$D_J^*(2650)$ & $2649.2\pm3.5\pm3.5$  & $140.2\pm17.1\pm18.6$ &LHCb \cite{Aaij:2013sza}&  $D^{*} \pi$ \\

$D_J(2740)$ & $2737.0\pm3.5\pm11.2$ & $73.2\pm13.4\pm25.0$ &LHCb \cite{Aaij:2013sza}&  $D^{*} \pi$ \\

$D_J^*(2760)$ & $2761.1\pm5.1\pm6.5$ & $74.4\pm3.4\pm37.0$ &LHCb \cite{Aaij:2013sza}&  $D^{(*)} \pi$ \\

$D_J(3000)$ & $2971.8\pm8.7$ & $188.1\pm44.8$ &LHCb \cite{Aaij:2013sza}&  $D^{*} \pi$ \\

$D_J^*(3000)$ & $3008.1\pm4.0$  & $110.5\pm11.5$ &LHCb \cite{Aaij:2013sza}&  $D \pi$ \\

\bottomrule[1pt]
\end{tabular}

\end{table*}
%%%%%%%%%%%%%%%%%%%%%%%%%%%%%%%%%%%%%% new %%%%%%%%%%%%%%%%%%%%%%%%%%%%%

In the past years, experimentalists have made great advances for observing charmed mesons and accordingly a list of charmed mesons collected in Particle Data Group (PDG) has become increasingly abundant \cite{Beringer:1900zz}. Candidates of
higher radial and orbital excitations in the charmed meson family
are the newly observed mesons $D(2550)$, $D^*(2600)$ $D^*(2760)$ and $D(2750)$ from BaBar \cite{delAmoSanchez:2010vq}, and $D_J(2580)$, $D^*_J(2650)$, $D^*_J(2760)$, $D_J(2740)$, $D_J(3000)$ and $D^*_J(3000)$ from LHCb \cite{Aaij:2013sza}, whose experimental information together with other observed charmed mesons is given in Table~\ref{table:review}.

This new observation and the present research status motivate us with great interest to carry out
systematic and phenomenological study of higher radial and orbital excitations in the charmed meson family, which will reveal the underlying properties of the observed charmed mesons and provide much more information for their further experimental search.

Although there have been a couple of works studying the heavy-light systems including charmed mesons with their decay modes \cite{Goity:1998jr,Di Pierro:2001uu,Bardeen:2003kt, Matsuki:1997da,Matsuki:2007zza,Matsuki:2006rz,Matsuki:2011xp,Ebert:1997nk},
in this work we focus mainly on application of our modified Godfrey-Isgur model developed in Ref. \cite{Song:2015nia} to calculating mass spectrum and decay behaviors of charmed mesons.
{Difference of our model from the Godfrey-Isgur model (GI) proposed in Ref. \cite{Godfrey:1985xj} is that the confining potential $br$ in the GI model is replaced by the screened one $V^{\mathrm{scr}}=b(1-e^{-\mu r})/\mu$ to reflect the unquenched effect (see Ref. \cite{Song:2015nia} for more details).}
% two important aspects of charmed mesons, i.e., their mass spectrum and decay behaviors.
%To study the mass spectrum of charmed mesons, we adopt the modified Godfrey-Isgur model developed in Ref. \cite{Song:2015nia}, which is different from the Godfrey Isgur model (GI) proposed in Ref. \cite{Godfrey:1985xj}.
%This work is a sequence of the modified Godfrey-Isgur model as an unquenched quark model \cite{Godfrey:1985xj}.
This model includes such a screening effect that
a linear confinement term $br$ be screened or softened at large distances by virtual quark pairs and dynamical fermions \cite{Laermann:1986pu,Born:1989iv}, which is important especially for higher radial and orbital excitations. Even though the GI model does not consider the screening effect, it has been successful to describe the low-lying states \cite{Godfrey:1985xj}. With the modified GI model, we will show a quite successful mass spectrum of the {high-lying} charmed meson family. Having the abundant experimental data, we are able to compare our theoretical values with the corresponding experimental results, which can not only test the reliability of our modified GI model, but also give some useful structure information of the observed charmed mesons. In addition, we will predict masses of some missing charmed mesons, which gives an important hint to experimentally explore these missing states.

One more valuable lesson we learned from the former work Ref. \cite{Song:2015nia} is that an experimental value of a mixing angle which is close to the one in the heavy quark limit determines relative decay widths, broad or narrow, with the same $J^P$ states in two spin multiplets. These results are found to be model-independent. The same scenario can be applied to the charmed meson family too and details will be described later.

After analyzing the mass spectrum of charmed mesons, in this work we further study two-body OZI-allowed decays of charmed mesons, where the quark pair creation (QPC) model \cite{Micu:1968mk,Le Yaouanc:1972ae,LeYaouanc:1988fx,vanBeveren:1979bd,vanBeveren:1982qb,Bonnaz:2001aj,roberts} is applied to calculating strong decays. With the modified GI model, we obtain the numerical wave functions of charmed mesons, which can be applied to calculating strong decays of charmed mesons, where we give partial and total widths of charmed mesons under discussion and some typical ratios relevant to these decays.
%The study of decay behaviors of charmed mesons can further reveal the underlying features of the charmed mesons under discussion.

Combining the analysis of mass spectrum with calculation of the decay widths, we can learn the properties of higher radial and orbital excitations in the charmed meson family. Furthermore, we can also shed light on the underlying structures of the observed charmed states under discussion, which is the main task of this work.

This paper is organized as follows. In Sect. \ref{sec2} we give a concise review on the observed charmed mesons. Then, the mass spectrum will be analyzed in Sect. \ref{sec3}, where the modified GI model is briefly introduced. In Sect. \ref{sec4} we will give a brief review of the QPC model and calculate the two-body OZI-allowed decay of the charmed mesons under discussion by the QPC model. This paper will end with a short summary in Sect. \ref{sec5}.

\section{Status of the observed charmed mesons}\label{sec2}

Before investigating the observed charmed mesons, we need to briefly review the present research status on candidates of higher radial and orbital excitations of the charmed meson family,

\subsection{$D^*_0(2400)$, $D_1(2430)$, $D_1(2420)$ and $D^*_2(2460)$}\label{subsec2}

As the first observed $P$-wave charmed meson, $D_1(2420)$ was reported in the $D^* \pi$ invariant mass distribution by the ARGUS Collaboration, where its measured mass and width are
$M=2420\pm6$ MeV and $\Gamma=70\pm21$ MeV, respectively \cite{Albrecht:1985as}. In 1989, the TPS Collaboration confirmed $D_1(2420)$ in the $D^{*+}\pi^-$ decay channel \cite{Anjos:1988uf}. According to this decay mode, its spin-parity %quantum number
is either $J^P=1^+$ or $2^+$.
The angular momentum analysis by the ARGUS Collaboration further shows that the observed $D_1(2420)$ has
$J^P=1^+$ \cite{Albrecht:1989pa}. The $D_1(2420)$ has been  confirmed by other experiments \cite{delAmoSanchez:2010vq,Albrecht:1988dj,Avery:1989ui,Frabetti:1993vv,Avery:1994yc,Ackerstaff:1997vc,
Abreu:1998vk,Abe:2003zm,Abe:2004sm,Abulencia:2005ry,Chekanov:2008ac,Aubert:2008zc,Abramowicz:2012ys} too.

The Belle Collaboration observed a broad state $D_1(2430)$ with mass $M=2427\pm26\pm20\pm15$MeV and width $\Gamma=384^{+107}_{-75}\pm24\pm70$ MeV by analyzing the $B\rightarrow D^* \pi \pi$ process, where its spin-parity is determined as $J^P=1^+$ by the helicity distributions \cite{Abe:2003zm}.
In 2006, the BaBar Collaboration studied the $D^*\pi$ invariant mass spectrum, where the broad $D_1(2430)$ was confirmed \cite{Aubert:2006zb}.

Besides $D_1(2430)$, Belle announced the observation of another broad state $D^*_0(2400)$, which has mass $M=2308\pm17\pm15\pm28$ MeV with $J^P=0^+$ \cite{Abe:2003zm}. The $D^*_0(2400)$ was also confirmed by the FOCUS and BaBar Collaborations with $M=2403\pm14\pm35$MeV \cite{Link:2003bd} and $M=2297\pm8\pm5\pm19$MeV \cite{Aubert:2009wg}, respectively.
Since different experiments gave quite different mass values for $D^*_0(2400)$, we will discuss mass dependence of the $D^*_0(2400)$ decay by varying mass in the range $(2290\sim 2350)$ MeV. The $D^*_0(2400)$ as a $P$-wave state with $J^P=0^+$ is supported by the theoretical works \cite{Godfrey:1986wj,Goity:1998jr,Di Pierro:2001uu,Matsuki:2007zza}.

The TPS Collaboration observed the charmed meson with $M=2459\pm3$MeV and $\Gamma=20\pm10\pm5$ in the invariant mass spectrum of $D^+\pi^-$  \cite{Anjos:1988uf}, which shows that
this meson has either $J^P=0^+$ or $2^+$. This observation was confirmed by the ARGUS Collaboration in the $D^+\pi^-$ channel, and their angular momentum analysis suggests the $J^P=2^+$ assignment to
this state \cite{Albrecht:1988dj}. Thus, this resonance is named $D_2^*(2460)$. Later, the CLEO Collaboration again confirmed the existence of $D_2^*(2460)$, where it has mass $M=2461\pm3\pm1$MeV and width $\Gamma=20^{+\ 9 + \ 9}_{-10 + 12} $ MeV. In addition, the ratio
\begin{eqnarray}
\frac{\mathcal{B}(D_2^*(2460)\rightarrow D^+\pi^-)}{\mathcal{B}(D_2^*(2460)\rightarrow D^{*+}\pi^-)}=2.3\pm0.8
\end{eqnarray}
was measured as well in Ref. \cite{Avery:1989ui}, which is consistent with the theoretical calculations in Refs. \cite{Rosner:1985dx,Godfrey:1986wj,Falk:1992cx,Eichten:1993ub,Goity:1998jr,Di Pierro:2001uu,Matsuki:2011xp}. In recent years, many other experiments have reported $D^*_2(2460)$ \cite{Albrecht:1989pa,Frabetti:1993vv,Avery:1994yc,Ackerstaff:1997vc,
Abreu:1998vk,Abe:2004sm,Abulencia:2005ry,Chekanov:2008ac,Aubert:2008zc,
delAmoSanchez:2010vq,Abramowicz:2012ys,Asratian:1995wx,Link:2003bd,Aubert:2009wg}.

The above experimental information indicates that there are two charmed mesons with $J^P=1^+$. In the heavy quark limit $m_Q\rightarrow\infty$, $\vec j_\ell=\vec S_q+\vec L$ is a good quantum number, where $\vec S_q$ is the spin of a light quark and $\vec L$ is its angular momentum. Thus, heavy-light mesons can be classified by $j_\ell^P$. Two $1S$-wave charmed mesons form one doublet $(0^-,1^-)$ with $j_l^P=\frac{1}{2}^-$.
Four $1P$-wave charmed mesons can be grouped into two doublets, $(0^+,1^+)$ and $(1^+,2^+)$, with $j_l^P=\frac{1}{2}^+$ and $j_l^P=\frac{3}{2}^+$, respectively.
Since states with $J^P=1^+$ in $(0^+,1^+)$ and $(1^+,2^+)$ doublets
decay into $D^*\pi$ via $S$-wave and $D$-wave
\cite{Godfrey:1986wj,Goity:1998jr,Di Pierro:2001uu,Matsuki:2007zza,Close:2005se,Godfrey:2005ww,Zhong:2008kd} in the heavy quark limit, respectively,
the charmed mesons with $J^P=1^+$ in $(0^+,1^+)$ and $(1^+,2^+)$ doublets have broad and narrow widths, respectively, which makes us
easily distinguish two $1^+$ charmed mesons experimentally observed, i.e., $D_1(2430)$ and $D_1(2420)$ belong to the doublets $(0^+,1^+)$ and $(1^+,2^+)$ \cite{Rosner:1985dx,Godfrey:1986wj,Eichten:1993ub,Goity:1998jr,Di Pierro:2001uu,Matsuki:2007zza}, respectively.

\subsection{$D(2550)$, $D_J(2580)$, $D^*(2600)$ and $D^*_J(2650)$}\label{subsec4}

The $D(2550)$ was observed by the BaBar Collaboration, which has the mass $M=2539.4\pm4.5\pm6.8$ MeV and width $\Gamma=130\pm12\pm13$ MeV \cite{delAmoSanchez:2010vq}. The $D(2550)$ is suggested to be a candidate for $D(2^1S_0)$ by the helicity distribution analysis \cite{delAmoSanchez:2010vq}. In 2013, an unnatural state $D_J(2580)$ was found in the $D^* \pi$ invariant mass spectrum by the LHCb Collaboration through the process $pp\rightarrow D \pi X$ \cite{Aaij:2013sza}. Since the resonance parameters of $D_J(2580)$ are similar to those of $D(2550)$, $D(2550)$ and $D_J(2580)$ are regarded as the same state.

The mass of $D(2550)$ or $D_J(2580)$ is consistent with the theoretical prediction of $D(2^1S_0)$ in Ref. \cite{Godfrey:1985xj}. In addition, the decay width of $D(2^1S_0)$ was calculated by the QPC model \cite{Close:2005se}, which is close to the lower limit of experimental width of $D(2550)/D_J(2580)$. In addition, the theoretical studies presented in Refs. \cite{Chen:2011rr,Lu:2014zua}
also show that $D(2550)$ can be $D(2^1S_0)$, the first radial excitation of $D$ mesons.
However, the authors in Ref. \cite{Di Pierro:2001uu,Sun:2010pg,Zhong:2010vq,Li:2010vx} indicated that the theoretical total width of $D(2^1S_0)$ is far below the experimental value of $D(2550)$.

The BaBar Collaboration reported a resonance $D^*(2600)$ with mass $M=2608.7\pm2.4\pm2.5$ MeV and width $\Gamma=93\pm6\pm13$ MeV, which is regarded as a radial excitation of $D^*$ by the helicity distribution analysis \cite{delAmoSanchez:2010vq} and they also measured the ratio \cite{delAmoSanchez:2010vq}
\begin{eqnarray}
\frac{\mathcal{B}(D^{*0}(2600)\rightarrow D^+\pi^-)}{\mathcal{B}(D^{*0}(2600)\rightarrow D^{*+}\pi^-)}=0.32\pm0.02\pm0.09. \label{26000}
\end{eqnarray}
A natural state $D^*_J(2650)$ was found in the $D^* \pi$ invariant mass spectrum by the LHCb collaboration through the process $pp\rightarrow D^* \pi X$ \cite{Aaij:2013sza}, where $D^*_J(2650)$ is tentatively identified as a $J^P=1^-$ state, a radial excitation of $D^*$. Therefore, $D^*_J(2650)$ \cite{Aaij:2013sza} and $D^*(2600)$ \cite{delAmoSanchez:2010vq} are the same state.

In Ref. \cite{Zeng:1994vj}, they predicted that mass of $D(2^3S_1)$ is 2620 MeV via the constituent quark model, which is in good agreement with the experimental value of $D^*(2600)$.
Moreover, the ratio $\Gamma(D(2^3S_1)^0\rightarrow D^+\pi^-) / \Gamma(D(2^3S_1)^0\rightarrow D^{*+}\pi^-)=0.47$ was predicted via the relativistic chiral quark model \cite{Goity:1998jr}, which is close to the upper bound of Eq. (\ref{26000}).
In Ref.\cite{Di Pierro:2001uu}, the authors calculated a mass spectrum and wave functions of charmed mesons via a relativistic quark model, and then adopted the obtained masses and wave functions as an input to estimate hadronic decay widths. Here, the predicted mass of $D(2^3S_1)$ is 2692 MeV which is heavier than $D^*(2600)$, while the predicted total width of $D(2^3S_1)$ is consistent with the experimental data of $D^*(2600)$.
Furthermore, the assignment $2^3S_1$ to $D^*(2600)$ is also supported by the studies in Refs. \cite{Sun:2010pg,Zhong:2010vq,Wang:2010ydc,Li:2010vx,Chen:2011rr,Yu:2014dda,Lu:2014zua}.

\subsection{$D^*(2760)$, $D^*_J(2760)$, $D(2750)$ and $D_J(2740)$}\label{subsec6}

The $D^*(2760)$ observed by the BaBar Collaboration in the $D \pi$ invariant mass spectrum \cite{delAmoSanchez:2010vq} can be assigned to a $D$-wave charmed meson since its mass is consistent with the theoretical prediction in Ref. \cite{Godfrey:1985xj}. Later, LHCb announced the observation of a natural state $D^*_J(2760)$ with mass $M=2761.1\pm5.1\pm6.5$ MeV and width $\Gamma=74.4\pm3.4\pm37.0$ MeV. %,  which can be identified as the $D(1^3D_1)$ state.
Both $D^*(2760)$ and $D^*_J(2760)$ can be regarded as the same state since these two states have the similar widths and masses \cite{Wang:2013tka}.

Comparison between the prediction in the Ref. \cite{Di Pierro:2001uu} and the experimental data of $D^*(2760)$ shows that $D^*(2760)$ can be either $D(1^3D_1)$ or $D(1^3D_3)$. However, the assignment of $D^*(2760)$
to $D(1^3D_1)$ or $D(1^3D_3)$ cannot be supported by the result shown in Ref. \cite{Close:2005se} since the calculated total widths of these two assignments are far larger than the experimental value. Later, in Ref. \cite{Sun:2010pg},
it was suggested that $D^*(2760)$ is a mixture of the $2^3S_1$ and $1^3D_1$ states, which is also supported by the study presented in Ref. \cite{Chen:2011rr}. However, calculation by the constituent quark model shows that the $D(1^3D_3)$ assignment of $D^*(2760)$ cannot be excluded \cite{Zhong:2010vq}, which is also supported by the works in Refs. \cite{Sun:2010pg,Wang:2010ydc,Li:2010vx,Colangelo:2012xi}.

Besides $D^*(2760)$, another state $D(2750)$ was also observed by the BaBar Collaboration in the $D^* \pi$ mass spectrum, where its mass and width are $M=2752.4\pm1.7\pm2.7$ MeV and $\Gamma=71\pm6\pm11$ MeV, respectively \cite{delAmoSanchez:2010vq}. Although $D(2750)$ can be a good candidate of a $D$-wave charmed meson according to the mass spectrum analysis in Ref. \cite{Godfrey:1985xj}, the helicity distribution analysis of $D(2750)$ does not support the $D(1^3D_1)$ and $D(1^3D_3)$ assignments \cite{delAmoSanchez:2010vq}. BaBar also gave
the ratio \cite{delAmoSanchez:2010vq}
\begin{eqnarray}
\frac{\mathcal{B}(D^*(2760)^0\rightarrow D^+\pi^-)}{\mathcal{B}(D(2750)^0\rightarrow D^{*+}\pi^-)}=0.42\pm0.05\pm0.11.
\end{eqnarray}

As an unnatural state, $D_J(2740)$ was found by the LHCb Collaboration, which has mass $M=2737.0\pm3.5\pm11.2$ MeV and $J^P=2^-$ \cite{Aaij:2013sza}. Due to the similarity between $D(2750)$ and $D_J(2740)$, it is possible that $D(2750)$ and $D_J(2740)$ are the same state. Before the observation of $D(2750)/D_J(2740)$, the masses of two $2^-$ charmed mesons were predicted in Ref. \cite{Di Pierro:2001uu}, where the masses of the $2^-$ charmed mesons belonging to the $(1^-,2^-)$ and $(2^-,3^-)$ doublets are
2883 MeV and 2775 MeV, respectively, which shows that $D(2750)/D_J(2740)$ as a $2^-$ state with $j_\ell^P=5/2^-$, i.e., $(2^-,3^-)$, is more favorable.
However, we also notice that the corresponding theoretical width of
this $2^-$ state is not consistent with the experimental value.
After observating $D(2750)/D_J(2740)$, the authors of Ref. \cite{Colangelo:2012xi} calculated the ratio $\Gamma(D^*(2760)^0\rightarrow D^+\pi^-)/ \Gamma(D(2750)^0\rightarrow D^{*+}\pi^-)$ by adopting an effective Lagrangian approach, which is consistent with the  experimental data, where $D^*(2760)$ and $D(2750)$ are identified  as the $1^3D_3$ and $2^-$ states in the $(2^-,3^-)$ doublet, respectively. The studies in Refs. \cite{Zhong:2010vq,Wang:2010ydc,Li:2010vx,Chen:2011rr} also suggested that $D(2750)$ is a $2^-$ state in the $(2^-,3^-)$ doublet.

\subsection{$D_J(3000)$ and $D^*_J(3000)$}\label{subsec9}

The LHCb Collaboration observed the unnatural state $D_J(3000)$ in the $D^* \pi$ invariant mass spectrum \cite{Aaij:2013sza} , where its resonance parameters are
\begin{eqnarray*}
M=2971.8\pm8.7 \,\mathrm{MeV},\quad
\Gamma=188.1\pm44.8 \,\mathrm{MeV}.
\end{eqnarray*}
Then, different theoretical groups carried out the study of $D_J(3000)$. In Refs. \cite{Sun:2013qca}, $D_J(3000)$ is regard as
the first radial excitation of $D_1(2430)$, which was also confirmed by Refs. \cite{Xiao:2014ura,Yu:2014dda}. However, other possible assignments to
$D_J(3000)$ were proposed, i.e., the $D(3^1S_0)$ \cite{Lu:2014zua} and $D(3^+)$ \cite{Yu:2014dda} assignments.

A natural state $D^*_J(3000)$ was also reported by LHCb in the $D \pi$ invariant mass spectrum \cite{Aaij:2013sza},
which has
\begin{eqnarray*}
M=3008.1\pm4.0 \,\mathrm{MeV},\quad
\Gamma=110.5\pm11.5 \,\mathrm{MeV} .
\end{eqnarray*}
The $D^*_J(3000)$ is variously explained as $D(2^3P_0)$ \cite{Sun:2013qca}, $D(1^3F_2)$ \cite{Yu:2014dda} and $D(1^3F_4)$ \cite{Xiao:2014ura,Yu:2014dda,Lu:2014zua}.
\\

%Until now, there have been so many observations of charmed mesons. It is a suitable time to carry out a systematical study of higher radial and orbital excitations in the charmed meson family by combining these experimental informations with theoretical results. In the following sections, we focus on this interesting research topic by performing the analysis of mass spectrum and calculation of the strong decay behaviors.

\section{mass spectrum}\label{sec3}

For the heavy-light meson system, we need to adopt a relativistic quark model to study their mass spectrum since a relativistic effect for a heavy-light meson system is significant. The Godfrey-Isgur (GI) model
can well describe the meson spectrum \cite{Godfrey:1985xj}, which is a typical quenched quark model. After the discovery of $D_{s0}(2317)$ \cite{Aubert:2003fg,Besson:2003cp,Abe:2003jk,Aubert:2006bk}, $D_{s1}(2460)$ \cite{Besson:2003cp,Abe:2003jk, Aubert:2003pe, Aubert:2006bk} and $X(3872)$ \cite{Choi:2003ue}, theorists realized that it is necessary to take into account coupled channel effects, especially for higher radial and orbital excitations of hadrons \cite{Eichten:2004uh,vanBeveren:2003kd,Dai:2006uz,Liu:2009uz}, where the coupled channel effects may change the meson spectrum. This motivates us to modify the GI model by considering the coupled channel effects.

In general, spontaneous creation of light quark-antiquark pairs inside a meson can soften a linear confinement potential $br$ by screening a color charge at distances larger than about one fermi \cite{Born:1989iv}, which is known as the screening effect. The screening effect has been seen by the unquenched Lattice QCD and holographic models \cite{Bali:2005fu,Armoni:2008jy,Bigazzi:2008gd}. The mass suppression can be caused by both the screening and coupled channel effects. Although the screening effect can be almost equivalent description of the coupled channel effect, we need to emphasize that the screening effect cannot depict the near-threshold effect as the coupled channel effect does \cite{Li:2009ad}, {which is obtained by studying charmonium spectrum. The authors of this paper made a comparison of the results by adopting the screening and coupled channel effects. Applying the idea in Ref. \cite{Li:2009ad} to our case, we can similarly study a charmed meson spectrum. However, it is a complicated task and can be assessed at a future theoretical work\footnote{We would like to thank the referee for his/her useful suggestion on this point}.}

In Refs. \cite{Mezoir:2008vx,Li:2009zu,Song:2015nia}, the screening effect was taken into account when studying the mass spectra of light mesons, charmonia and charmed-strange mesons. Mezzoir {\it et al.} provide description of a highly excited light-quark meson spectrum by flattening the confining potential $br$ at distances larger than $r_s$ \cite{Mezoir:2008vx}. The screened potential model \cite{Chao:1992et,Ding:1993uy} was adopted to compute the charmonium spectrum \cite{Li:2009zu}. In our recent work \cite{Song:2015nia}, the screening effect was introduced to modify the GI model, where the mass spectrum of charmed-strange meson family with this treatment is greatly improved compared with the results of the GI model. As a sister work of Ref. \cite{Song:2015nia}, the present work focuses on the charmed mesons applying the modified GI model \cite{Song:2015nia} to obtain their mass spectrum.

{In the GI model, the confining potential $br$ is smeared out to include relativistic effects, i.e.,
\begin{eqnarray}
\tilde V(r)= \int d^3 \bm{r}^\prime
\rho_{12}(\bm{r-r^\prime})br^\prime.
\end{eqnarray}
Here, the GI model is a quenched quark model in the sense that the effect of quark-antiquark pairs is not introduced \cite{Godfrey:1985xj}.}
In order to take account of the screening effect in the GI model, the confining potential $br$ is replaced with \cite{Chao:1992et,Ding:1993uy}
\begin{eqnarray}
br\to V^{\mathrm{scr}}(r)=\frac{b(1-e^{-\mu r})}{\mu},
\end{eqnarray}
where $V^{\mathrm{scr}}(r)$ behaves like $br$ at short distances and constant ${b}/{\mu}$ at large distances. Furthermore, the smearing function is introduced to take into account nonlocality property of potentials \cite{Godfrey:1985xj}, i.e.,
\begin{eqnarray}
\tilde V^{\mathrm{scr}}(r)&=& \int d^3 \bm{r}^\prime
\rho_{12}(\bm{r-r^\prime})\frac{b(1-e^{-\mu r^\prime})}{\mu}.\label{5}
\end{eqnarray}
The detailed explanation of how to introduce the screening effect into the GI model can be found in Ref. \cite{Song:2015nia}.

\renewcommand{\arraystretch}{1.3}
\begin{table}[htbp]
\caption{The calculated masses of charmed mesons by the modified GI model and comparison with those obtained by the GI model. Here, we take several $\mu$ values, $\mu=0.01\, , 0.02 \,, 0.03 \, ,$ and $0.04 \, \mathrm{GeV}$ to show $\mu$ dependence of the modified GI model. Values in brackets for the GI model and $\mu=0.03$ are those of $R=1/\beta$, which can be determined by solving $\int\Psi^{\mathrm{SHO}}_{nLM}(\mathbf{p})^2 p^2 d^3\mathbf{p}=\int\Phi(\mathbf{p})^2 p^2 d^3\mathbf{p}$, where $\Psi^{\mathrm{SHO}}_{nLM}(\mathbf{p})$ is an SHO wave function and $\Phi(\mathbf{p})$ is the wave function of a charmed meson which we obtain by solving an eigenvalue equaiton. We need to emphasize that we do not consider mixing among states with the same quantum number when presenting the results. $\mu$ is in units of GeV, while $R$ is in units of GeV$^{-1}$. }
\centering
\begin{tabular}{l c | c c c c  }\toprule[1pt]

&GI model  &\multicolumn{4}{c}{Modified GI model}
\\&&$\mu=0.01$& $\mu=0.02$ &$\mu=0.03$&$\mu=0.04$\\ \midrule[1pt]
${1}^1S_0$&1874(1.52)&1869&1865&1861(1.54)&1855\\
${2}^1S_0$&2583(2.08)&2566&2550&2534(2.22)&2518\\
${3}^1S_0$&3068(2.33)&3037&3005&2976(2.50)&2945\\
${1}^3S_1$&2038(1.85)&2032&2027&2020(1.89)&2015\\
${2}^3S_1$&2645(2.17)&2628&2610&2593(2.33)&2576\\
${3}^3S_1$&3111(2.38)&3079&3047&3015(2.56)&2983\\
${1}^1P_1$&2457(2.00)&2447&2436&2426(2.08)&2415\\
${2}^1P_1$&2933(2.27)&2909&2885&2861(2.44)&2837\\
${1}^3P_0$&2398(1.85)&2387&2376&2365(1.92)&2354\\
${2}^3P_0$&2932(2.22)&2907&2881&2856(2.38)&2831\\
${1}^3P_1$&2465(2.00)&2453&2441&2431(2.08)&2419\\
${2}^3P_1$&2952(2.27)&2927&2902&2877(2.44)&2852\\
${1}^3P_2$&2501(2.22)&2490&2479&2468(2.33)&2456\\
${2}^3P_2$&2957(2.38)&2933&2908&2884(2.56)&2859\\
${1}^1D_2$&2827(2.27)&2807&2791&2773(2.38)&2755\\
${2}^1D_2$&3225(2.44)&3193&3160&3128(2.63)&3095\\
${1}^3D_1$&2816(2.13)&2798&2780&2762(2.27)&2744\\
${2}^3D_1$&3231(2.33)&3198&3164&3131(2.56)&3097\\
${1}^3D_2$&2834(2.27)&2816&2797&2779(2.38)&2761\\
${2}^3D_2$&3235(2.44)&3202&3169&3136(2.63)&3102\\
${1}^3D_3$&2833(2.38)&2815&2797&2779(2.56)&2761\\
${2}^3D_3$&3226(2.50)&3194&3162&3129(2.70)&3097\\
${1}^1F_3$&3123(2.44)&3097&3072&3046(2.63)&3019\\
${1}^3F_2$&3132(2.33)&3106&3080&3053(2.50)&3027 \\
${1}^3F_3$&3129(2.44)&3104&3078&3051(2.63)&3025\\
${1}^3F_4$&3113(2.50)&3088&3063&3037(2.70)&3011\\  \bottomrule[1pt]
\end {tabular}\\
\label{table:spectrum} % is used to refer this table in the text
\end{table}

\renewcommand{\arraystretch}{1.6}
\begin{table*}[htbp]
\caption{\label{table:comparision}Comparison of experimental data and theoretical results. Here, we also list the $\chi^2$ values for different models. The notation $L_L$ is introduced to express mixing states of $^1L_L$ and $^3L_L$. Here, the results listed in the last column are calculated by the modified GI model with $\mu=0.03 $ GeV which gives the least $\chi^2$ value among several $\mu$'s.}
\centering
\begin{tabular}{ c c c c c c }\toprule[1pt]
&$n \ ^{2S+1}L_J$ &Experimental values \cite{Beringer:1900zz} & GI model \cite{Godfrey:1985xj}& Modified GI model \\ \midrule[1pt]
$D$ &$1 \ ^1S_0$&$1864.84\pm0.07$&1874&1861 \\
$D^{\ast}$&$1 \ ^3S_1$&$2010.26\pm0.07$&2038& 2020 \\
$D_{0}^\ast(2400)$&$1 \ ^3P_0$&$2318\pm29$&2398&2365 \\
$D_{1}(2420)$&$1 \ P_1$&$2421.4\pm0.6$&2467&2434 \\
$D_{1}(2430)$&$1 \ P_1$&$2427\pm26\pm25$&2455&2424 \\
$D_{2}^\ast(2460)$&$1 \ ^3P_2$&$2464.3\pm1.6$&2501& 2468\\
$D(2550)$ &$2 \ ^1S_0$ &$2539.4\pm4.5\pm6.8$  \cite{delAmoSanchez:2010vq} &2583&2534\\
$D^*(2600)$ & $2 \ ^3S_1$&$2608.7\pm2.4\pm2.5$  \cite{delAmoSanchez:2010vq} &2645 &2593\\
$D(2750)$ & $1 \ D_2$&$2752.4\pm1.7\pm2.7$  \cite{delAmoSanchez:2010vq} & 2845&2789\\
$D^*(2760)$ &$1 \ ^3D_3$& $2763.3\pm2.3\pm2.3$  \cite{delAmoSanchez:2010vq} &2833&2779
%\\ &{\color{red}$1 \ ^3D_1$}&  &{\color{red}2820}&{\color{red}2762}
%$D_J(3000)$ &$3 \ ^1S_0$& $2971.8\pm8.7$  \cite{Aaij:2013sza}&3068&2976\\
%$D_J^*(3000)$ &$3 \ ^3S_1$& $3008.1\pm4.0$  \cite{Aaij:2013sza} & 3111&3015
\\ \midrule[1pt]
$\chi^2$&--&--&{45677}&{5748 }\\ \bottomrule[1pt]
\end {tabular}
\label{table:compare}
\end{table*}

%\iffalse
\begin{figure*}[htpb]
\includegraphics[width=0.9\textwidth]{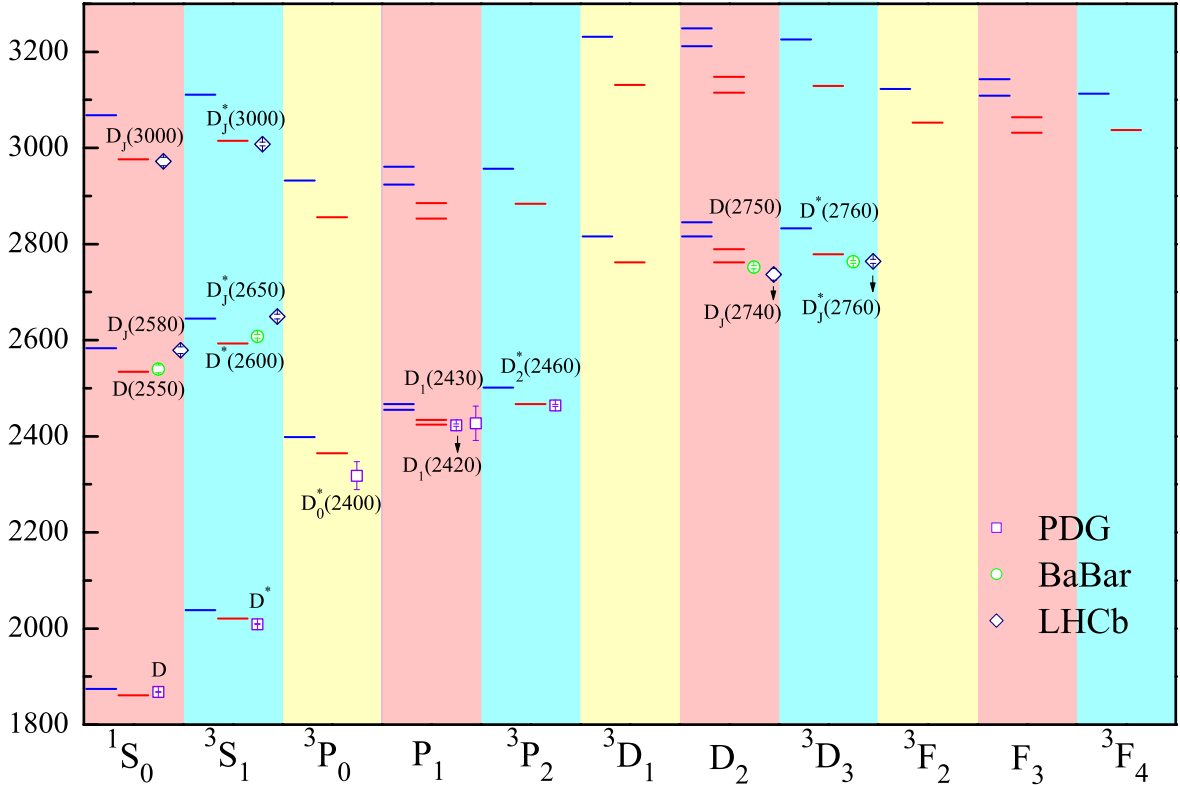}
\caption{\label{fig:spectrum} (color online). Mass spectrum of charmed mesons (in units of MeV).
Here, the blue lines stand for the results of the GI model \cite{Godfrey:1985xj}, while the red lines are those of the modified GI model with $\mu=0.03$ GeV. The purple squares denote the experimental data taken from PDG \cite{Beringer:1900zz}, while the green circles and blue lozenges are experimental masses of Ref. \cite{delAmoSanchez:2010vq} and Ref. \cite{Aaij:2013sza}, respectively. The corresponding $^{2S+1}L_J$ quantum numbers are listed on the abscissa. In addition, when there exists
a mixture of $n^1L_L$ and $n^3L_L$ states, we use a notation $L_L$.}
\end{figure*}
%\fi

The mass spectrum of charmed mesons is calculated in the GI model and the modified GI model, which is shown in Table \ref{table:spectrum}. Here, all the parameters of the modified GI model are the same as those of the GI model \cite{Godfrey:1985xj} except for the additional parameter $\mu$ in $\tilde V^{\mathrm{scr}}(r)$. {In this Table, we present the theoretical predictions of the mass spectrum with several values of $\mu$, $0.01\, \mathrm{GeV}, 0.02\, \mathrm{GeV}, 0.03  \,\mathrm{GeV},$ and $0.04 \, \mathrm{GeV}$. In order to choose a suitable value for $\mu$, we calculate the $\chi^2$ of predicted masses for each $\mu$. In Table \ref{table:compare} we list the theoretical results with the least value of $\mu=0.03$ GeV together with experimental data. Here the concrete expression for $\chi^2$ is,
\begin{eqnarray}
\chi^2=\sum_i\left(\frac{\mathcal{A}_{\mathrm{Th}}(i)-\mathcal{A}_{\mathrm{Exp}}(i)}{\mathrm{Error}(i)}\right)^2,
\end{eqnarray}
where $\mathcal{A}_{\mathrm{Th}}(i)$ and $\mathcal{A}_{\mathrm{Exp}}(i)$ are theoretical and experimental values, respectively.  $\mathrm{Error}(i)$ denotes the experimental error of the mass of a charmed meson. A minimum of $\chi^2$ is obtained by our model with $\mu=0.03$ GeV. Eventually, the modified GI model with this value of $\mu$ improves the whole description of the charmed meson mass spectrum compared with the GI model (see Table \ref{table:compare} for more details).}

In Table \ref{table:comparision}, we further make a comparison between experimental data and theoretical values obtained via the modified GI model. Besides the $2^1S_0$ and $2^3S_1$ states, a difference between theoretical and experimental values of other charmed mesons is less than 20 MeV. We can conclude from the mass spectrum analysis, i.e.,

\begin{enumerate}

\item{Two $1S$ states and four $1P$ states in the charmed meson family can be well reproduced by the modified GI model.}

\item Both $D(2550)$ reported by the BaBar Collaboration \cite{delAmoSanchez:2010vq} and $D_J(2580)$ from the LHCb Collaboration \cite{Aaij:2013sza} are usually considered as a candidate of $D(2^1S_0)$. Here, the mass of $D(2550)$ is quite close to the theoretical mass of $D(2^1S_0)$, while the mass of $D_J(2580)$ is larger than the theoretical mass of $D(2^1S_0)$ by about 40 MeV.

\item There exist two experimental results $D^*(2600)$ and
$D^*_J(2650)$ from BaBar \cite{delAmoSanchez:2010vq} and LHCb \cite{Aaij:2013sza}, both of which can be a candidate of $D(2^3S_1)$. Our result shows that $D^*(2600)$ is closer to the theoretical mass of $D(2^3S_1)$.

\item{$D(2750)/D_J(2740)$ is a good candidate of $D(1D_2)$, while $D^*(2760)/D^*_J(2760)$ corresponds to  $D(1^3D_1)$ or $D(1^3D_3)$.}

\item{$D_J(3000)$ and $D_J^*(3000)$ can be candidates of $D(3^1S_0)$ and $D(3^3S_1)$, respectively. In addition, the mass spectrum study cannot exclude a possibility of $D_J(3000)$ and $D_J^*(3000)$ as the $1F$ states in the charmed meson family. We also notice that the theoretical masses of the $2P$ states  are about 100 MeV smaller than the corresponding experimental data of $D_J(3000)$ and $D_J^*(3000)$ .
    }

\end{enumerate}

{We also notice a study of the heavy-light meson spectroscopy within the framework of the QCD-motivated relativistic quark model
based on the quasipotential approach \cite{Ebert:2009ua}. The conclusion of mass spectrum obtained above in the present work is consistent with those of Ref. \cite{Ebert:2009ua}.}

In order to clearly identify the properties of the observed charmed mesons, we need to perform a systematic study of their decay behaviors, which is the main task in the next section.

\section{Strong decay behaviors}\label{sec4}

Study of  the two-body Okubo-Zweig-Iizuka (OZI) allowed strong decay behaviors of the observed and the predicted charmed mesons can provide much more information on the features of the charmed mesons under discussion, which includes total and partial decay widths. What is more important is that the inner structure of the observed charmed mesons given by the mass spectrum analysis in Sect. \ref{sec3} can be further tested here.

As an effective approach to study the OZI allowed strong decays of hadrons, the quark pair creation (QPC) model is applied to compute the OZI-allowed strong decays of charmed mesons, which was first proposed by Micu \cite{Micu:1968mk} and was further developed by the Orsay group \cite{Le Yaouanc:1972ae,LeYaouanc:1988fx,vanBeveren:1979bd,vanBeveren:1982qb,Bonnaz:2001aj,roberts}. Here, a meson decay occurs through a flavor and color singlet quark-antiquark pair created from the vacuum. To depict a $q\bar{q}$ pair creation from the vacuum, the transition operator $\mathcal{T}$ is introduced as,
\begin{eqnarray}
\mathcal{T}&=&-3\gamma \sum \limits_{m} \langle1m;1-m|00\rangle \int d\mathbf{p}_3 d\mathbf{p}_4 \delta^3(\mathbf{p}_3+\mathbf{p}_4)\nonumber\\
&&\times \mathcal{Y}_{1m}\left(\frac{\mathbf{p}_3-\mathbf{p}_4}{2}\right) \chi^{34}_{1,-m} \phi^{34}_{0} \left(\omega^{34}_{0}\right)_{ij}b^{\dag}_{3i}(\mathbf{p}_3) d^{\dag}_{4j}(\mathbf{p}_4), \label{eq:transition}
\end{eqnarray}
where $\gamma$ is a dimensionless constant which reflects the strength of creating a quark-antiquark pair from the vacuum.
{In Ref. \cite{Ye:2012gu}, $\gamma=8.7$ is obtained for the $u\overline{u}/d\overline{d}$ pair creation by fitting with the experimental data, where the $SU(3)$ flavor singlet wave function was adopted, while for reflecting the $SU(3)$ breaking effect we take
$\gamma=8.7/\sqrt{3}$ for the  $s\overline{s}$ pair creation as suggested in Ref. \cite{Le Yaouanc:1972ae}}.
We need to explain why there exists an extra $1/\sqrt{3}$ factor for the $s\overline{s}$ pair creation.
In Ref. \cite{Le Yaouanc:1972ae}, the authors realized the $SU(3)$ and $SU(4)$ breaking by defining the flavor function $\phi_0$ of a $q\bar q$ pair created from the vacuum, i.e., $\phi_0$ reads
$\phi_0=u \bar{u}+d \bar{d}+\sigma_s(s \bar{s})+\sigma_c(c \bar{c})$ where $\sigma_s=m_u/m_s$, $\sigma_c=m_u/m_c$, and
$\sigma_s $ expresses the $SU(3)$ breaking.
In oder to check whether the estimate of $\sigma_s$ is in agreement with
hadron spectroscopy, the authors of Ref. \cite{Le Yaouanc:1972ae} calculated ratios
$r_{a_2(1320)} = \Gamma(a_2(1320) \to \pi \eta)/ \Gamma(a_2(1320) \to K\bar{K})$ and  $r_{f_2(1270)} = \Gamma(f_2(1270)\to \pi \pi)/\Gamma(f_2(1270) \to  K\bar{K})$
by the QPC model to obtain $r_{a_2}=0.8/(\sigma_s)^2$ and $r_{f}=9.6/(\sigma_s)^2$.
Comparing the above ratios with the experimental data $r_{a_2}=3.2$ and $r_{f}=30$,
they found that $\sigma_s \approx 1/\sqrt{3}$ is a reasonable estimate.
In our work we adopt $\phi_0^{34}=(u \bar u+d \bar d+s \bar s)/\sqrt{3}$ as the flavor $SU(3)$ singlet wave function. To show the $SU(3)$ breaking,
the creation strength for $u \bar u$ and $d \bar{d}$ is defined as $\gamma$, while for the strength of $s \bar{s}$ creation we have $\gamma/\sqrt{3}$, which is equivalent to the description in Ref. \cite{Le Yaouanc:1972ae}. In Eq.~(\ref{eq:transition}), symbols $\mathbf{p}_3$  and  $\mathbf{p}_4$ stand for momenta of quark and antiquark, respectively. $\mathcal{Y}_{\ell m}(\mathbf{p})=|\mathbf{p}|^\ell Y_{\ell m}(\mathbf{p})$ is the solid harmonic polynomial and $\chi^{34}_{1,-m}$ is the spin triplet state. The quantum number of a quark-antiquark pair is $J^{PC}=0^{++}$ determined by coupling the orbital angular momentum with the spin angular momenta, which indicates the conservation of angular momentum $J$, $P$ parity and $C$ parity in the course of strong interaction. $\phi^{34}_{0}=(u \overline{u}+d \overline{d}+s  \overline{s})/\sqrt{3}$ and $\left(\omega^{34}_{0}\right)_{ij}=\delta_{ij}/\sqrt{3}$ are the $SU(3)$ flavor and color functions, respectively, with $i$ and $j$ being the color indices. %\footnote{{\color{red}There are four reasons why we do not consider the $SU(4)$ flavor singlet function : 1)	 If you check the published papers relevant to the QPC model, you can find that the $SU(3)$ flavor singlet $\phi_0 = (u \bar u + d \bar d + s \bar s)/\sqrt{3}$ has been extensively adopted till now \cite{Blundell:1996as,Blundell:1995ev}.
%2) Even if we take the $SU(4)$ flavor singlet expression, the present results and conclusions remain the same. The difference between the $SU(3)$ and the $SU(4)$ flavor singlet functions is in their overall normalization coefficients since the creation strength $\gamma$ from the vacuum is determined by the experimental data. If we adopt the new $SU(4)$ flavor singlet flavor function, we replace only the $\gamma$ value obtained from the $SU(3)$ flavor singlet by newly defined $\gamma\prime=2\gamma/\sqrt{3}$ in the calculation.
%3) In Ref.  \cite{Le Yaouanc:1972ae}, the authors concluded that the creation strength of $c\bar{c}$ from the vacuum is suppressed compared with the creation strength of $u\bar{u}/d\bar{d}$ and $s\bar{s}$ from the vacuum.
%4) In addition, the threshold of $J/\psi+D$ is about 5 GeV far larger than all the observed charmed mesons.}}.

The transition matrix of a process $A \rightarrow BC$ can be expressed as
\begin{eqnarray}
\langle BC|T|A\rangle=\delta^3(\mathbf{p}_B+\mathbf{p}_C)\mathcal{M}^{M_{JA}{M_{JB}}{M_{JC}}},
\end{eqnarray}
where $\mathbf{p}_B$ and $\mathbf{p}_C$ are the momenta of mesons $B$ and $C$, respectively. $|A\rangle$, $|B\rangle$ and $|C\rangle$ denote mock states \cite{Hayne:1981zy}. The mock state of a meson $A$ can be defined as
\begin{eqnarray}
&&|A(n^{2S+1}L_{JM_J})(\mathbf{p}_A)\rangle\nonumber\\
&&=\sqrt{2E}\sum\limits_{ {M_{S}},{M_{L}}}\langle LM_L;SM_{S}|JM_{J}\rangle\chi^{A}_{S,M_S} \nonumber\\
&&\quad\times \phi^A \omega^A\int d\mathbf{p}_1 d\mathbf{p}_2 \delta^3(
\mathbf{p}_A-\mathbf{p}_1-\mathbf{p}_2) \nonumber\\ &&\quad\times\Psi^A_{nLM_L}(\mathbf{p}_1,\mathbf{p}_2)|q_1(\mathbf{p}_1)\bar{q}_2(\mathbf{p}_2)\rangle,
\end{eqnarray}
where $\chi^{A}_{S,M_S}$, $\phi^A$ and $\omega^A$ are spin, flavor and color wave functions of a meson $A$, respectively.
$\Psi^A_{nLM_L}(\mathbf{p}_1,\mathbf{p}_2)$ is a spatial wave function of meson $A$, which can be obtained in the modified GI model. Furthermore, the amplitude ${\mathcal{M}}^{M_{JA}{M_{JB}}{M_{JC}}}$ can be related to the partial wave amplitude $\mathcal{M}^{JL}$ via the Jacob-Wick formula \cite{Jacob:1959at}, i.e.,
\begin{eqnarray}
\mathcal{M}^{JL}(A\rightarrow BC)&=&\frac{\sqrt{2L+1}}{2J_A+1}\sum \limits_{ {M_{JB}},{M_{JC}}}\langle L0;JM_{JA}|J_AM_{JA}\rangle\nonumber\\&&\times
\langle J_BM_{JB};J_CM_{JC}|JM_{JA}\rangle \mathcal{M}^{M_{JA}{M_{JB}}{M_{JC}}}.\nonumber
\end{eqnarray}
Therefore, the total decay width can be expressed as
\begin{eqnarray}
\Gamma=\pi^2\frac{|\mathbf{p}_B|}{m_A^2}\sum \limits_{J,L}|\mathcal{M}^{JL}|^2.
\end{eqnarray}

After this brief introduction of the QPC model, we perform a phenomenological analysis of charmed mesons in the following.
When calculating a decay width, we adopt the numerical wave function for a charmed meson obtained in this work and the one for charmed-strange meson from Ref. \cite{Song:2015nia}.  Additionally, we still employ
the simple harmonic oscillator wave function for light mesons such as $\pi$ and $K$, where the corresponding $\beta$ values are taken from Ref. \cite{Godfrey:1986wj}.
We need to emphasize that the mass is taken from PDG \cite{Beringer:1900zz} for the observed meson. For the charmed mesons which are still missing, we use the theoretical predictions calculated in the modified GI model, the results listed in Table. \ref{table:spectrum} and/or Fig. \ref{fig:spectrum}, as an input.

\subsection{$1P$ states}

\begin{figure}
\includegraphics[width=0.49\textwidth]{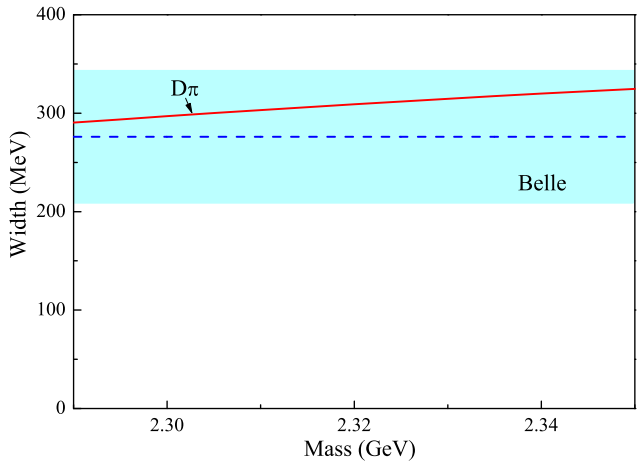}
\caption{\label{fig:d2400}Mass dependence of the width of $D_{0}^*(2400)$ and comparison with the Belle data \cite{Abe:2003zm}. The mass range of $D_{0}^*(2400)$ is $2920\sim 2350$ MeV.}
\end{figure}

As the $1^3P_0$ state, $D_{0}^*(2400)$ has been observed by three different experiments. However, the experimental masses are quite different from each other as shown in Table \ref{table:review}. Therefore, in this work we take the mass range ($2.29\sim 2.35$ GeV) of $D_{0}^*(2400)$ to discuss mass dependence of the calculated decay width. Here, $D_{0}^*(2400)$ only decays into $D \pi$. In Fig. \ref{fig:d2400}, we present the mass dependence of the decay width of $D_{0}^*(2400)$. We find that our results are consistent with experimental data $\Gamma=276\pm21\pm63$ \cite{Abe:2003zm}, $\Gamma=240\pm55\pm59$ \cite{Link:2003bd}, and $\Gamma=273\pm12\pm48$ \cite{Aubert:2009wg}.

%36.3-45.4%

In the following, we study $D_{1}(2420)$ and $D_{1}(2430)$, which are
mixtures of the $1^1P_1$ and $1^3P_1$ states. $D_{1}(2420)$ and $D_{1}(2430)$ satisfy the relation
\begin{equation}
 \left(
  \begin{array}{c}
   |D_{1}(2430)\rangle\\
   |D_{1}(2420)\rangle\\
  \end{array}
\right )=
\left(
  \begin{array}{cc}
    \cos\theta_{1P} & \sin\theta_{1P} \\
   -\sin\theta_{1P} & \cos\theta_{1P}\\
  \end{array}
\right)
\left(
  \begin{array}{c}
    |1^1P_1  \rangle \\
   |1^3P_1 \rangle\\
  \end{array}
\right),\label{m1}
\end{equation}
where the mixing angle is $\theta_{1P}=-54.7^\circ$ which is determined by the heavy quark limit \cite{Godfrey:1986wj,Matsuki:2010zy, Barnes:2002mu}.

\begin{figure}[!htbp]
\includegraphics[width=0.49\textwidth]{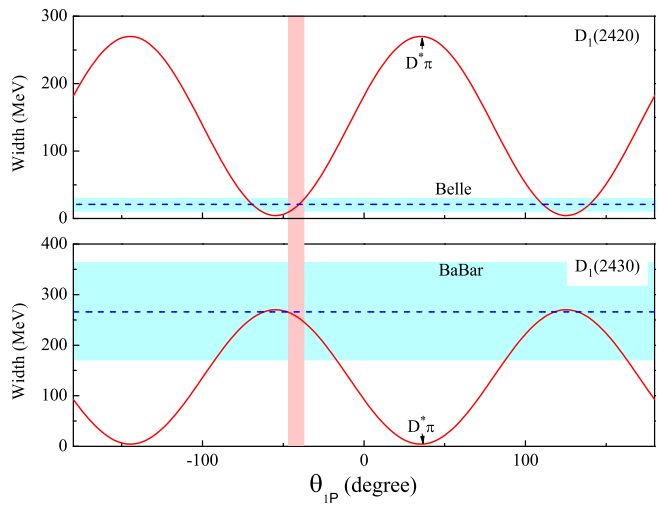}
\caption{\label{fig:1p}The $\theta_{1P}$ dependence of the decay widths  of $D_{1}(2420)$ and $D_{1}(2430)$ .}
\end{figure}

The width of $D_{1}(2420)$ ($\Gamma=70\pm21$) was first measured by the ARGUS Collaboration \cite{Albrecht:1985as}. However, different experiments provided  different widths as $\Gamma=13\pm6^{+10}_{-5}$ MeV \cite{Albrecht:1989pa}, $\Gamma=23^{+8+10}_{-6 -3}$ MeV \cite{Avery:1989ui}, and $\Gamma=21\pm5\pm8$ MeV \cite{Abe:2004sm}. Here, we take the median as the width, i.e., the measurement $\Gamma=21\pm5\pm8$ MeV from the Belle Collaboration \cite{Abe:2004sm}. $D_{1}(2430)$ is a broad state, which has the width $\Gamma=384^{+107}_{-75}\pm75$ MeV given by Belle \cite{Abe:2003zm}  and $\Gamma=266\pm96$ MeV by BaBar \cite{Aubert:2006zb}.
Both $D_{1}(2430)$ and $D_{1}(2420)$ only decay into $D\pi$.
In Fig. \ref{fig:1p}, we show the decay widths of $D_{1}(2420)$ and $D_{1}(2430)$ depending on the mixing angle $\theta_{1P}$, where
our results are consistent with the experimental data when taking $-45.6^\circ<\theta_{1P}<-37.2^\circ$, which is close to $\theta_{1P}=-54.7^\circ=-\arcsin(\sqrt{2/3})$ in the heavy quark limit \cite{Godfrey:1986wj,Matsuki:2010zy, Barnes:2002mu}.

The $D_2^*(2460)$ is considered as the $1^3P_2$ state. Its decay channels are $D\pi$, $D^*\pi$, $D\eta$ and $D_sK$, whose theoretical values of decay widths are shown in Table \ref{channels}. The total width is 14.86 MeV and the branching ratio $\mathcal{B}(D_2^*(2460)^0\rightarrow D^+\pi^-)/\mathcal{B}(D_2^*(2460)^0\rightarrow D^{*+}\pi^-)$ is 1.96.
There are several different experimental widths for $D_2^*(2460)$, i.e., $\Gamma=20\pm10\pm5 $ MeV from the TPS Collaboration\cite{Anjos:1988uf}, $\Gamma=20^{+\ 9 + \ 9}_{-10 - 12} $ MeV from the CLEO Collaboration \cite{Avery:1989ui} and $45.6\pm4.4\pm6.7 $ MeV from the Belle Collaboration \cite{Abe:2003zm}. Our theoretical width is consistent with the TPS data \cite{Anjos:1988uf} and the CLEO data \cite{Avery:1989ui}. Furthermore, our ratio is in good agreement with experimental values
$\mathcal{B}( D^+\pi^-)/\mathcal{B} (D^{*+}\pi^-)=2.3\pm0.8$ \cite{Avery:1989ui} and
$\mathcal{B}( D^+\pi^-)/\mathcal{B}( D^{*+}\pi^-)=1.9\pm0.5$ \cite{Abe:2003zm} measured by CLEO and Belle, respectively.

In this subsection, $D_{0}^*(2400)$, $D_{1}(2420)$, $D_{1}(2430)$ and $D_2^*(2460)$ which are well established as $1P$ charmed mesons have been analyzed, which makes us safely adopt the QPC model and the related model parameters to further study the following charmed mesons under discussion.

\renewcommand{\arraystretch}{1.4}
 \begin{table} \centering
 \caption{The partial and total widths of charmed mesons $D_2^*(2460)$, $D(2550)/D_J(2580)$, and $D(2600)/D^*_J(2650)$. We mark the OZI-forbidden channels by ``--'', If the channels depend on other parameters, we label them with ``$\Box$'' and show the widths in Fig. \ref{fig:2s1d}. All values are in units of MeV.}\label{channels}
\begin{tabular}{lcccccc}
 \\   \toprule[1pt]
Channels       &$D_2^*(2460)$&$D(2550)/D_J(2580)$&$D(2600)/D^*_J(2650)$\\ \midrule[1pt]
$D\pi$          &9.17               &--    &$\Box$  \\
$D\eta$         &0.02               &--    &$\Box$  \\
$D_sK$          &$1.0\times 10^{-4}$&--    &$\Box$  \\
$D^*\pi$        &4.67               &67.56 &$\Box$  \\
$D^*\eta$       &--                 &--    &$\Box$   \\
$D_s^*K$        &--                 &--    &$\Box$  \\
$D^*_0(2400)\pi$&--                 &4.09  & --     \\
$D^*_2(2460)\pi$&--                 &--    &$\Box$  \\
$D_1(2420)\pi$  &--                 &--    &$\Box$  \\
$D_1(2430)\pi$  &--                 &--    &$\Box$  \\ \midrule[1pt]
Total width     &14.86              &71.65 &--      \\  \bottomrule[1pt]
\end{tabular}
\end{table}

\subsection{$2S$ and $1D$ states}\label{m1}

The $D(2550)$/$D_J(2580)$ \cite{delAmoSanchez:2010vq,Aaij:2013sza} is usually considered as a candidate of the $2^1S_0$ state. The main decay channels of $D(2550)$ are $D^*\pi$ and $D^*_0(2400)\pi$ as shown in Table \ref{channels}, which can explain why BaBar and LHCb first observed $D(2550)$/$D_J(2580)$ in the $D^*\pi$ channel. The total width is obtained as 71.65 MeV which is comparable with the lower bound of the BaBar data \cite{delAmoSanchez:2010vq} and is smaller than the LHCb value \cite{Aaij:2013sza}. Considering this situation, we also suggest to do more precise measurement of the resonance parameters of $D(2550)$/$D_J(2580)$, which will be helpful for further testing the $2^1S_0$ assignment to $D(2550)$/$D_J(2580)$.

In the following, we study $D^*(2600)/D^*_J(2650)$ \cite{Aaij:2013sza,delAmoSanchez:2010vq} with $J^P=1^-$, which is a mixture of the $2^3S_1$ and $1^3D_1$ states.
Here, $D^*(2600)$ and its orthogonal partner satisfy
\begin{equation}
 \left(
  \begin{array}{c}
   |D^*(2600)\rangle\\
   |D^{*\prime}(1^-)\rangle\\
  \end{array}
\right )=
\left(
  \begin{array}{cc}
    \cos\theta_{SD} & \sin\theta_{SD} \\
   -\sin\theta_{SD} & \cos\theta_{SD}\\
  \end{array}
\right)
\left(
  \begin{array}{c}
    |2^3S_1  \rangle \\
   |1^3D_1 \rangle\\
  \end{array}
\right),\label{sd}
\end{equation}
where the mixing angle $\theta_{SD}$ is introduced to describe mixing between $D(2^3S_1)$ and $D(1^3D_1)$.

\begin{figure}
\includegraphics[width=0.49\textwidth]{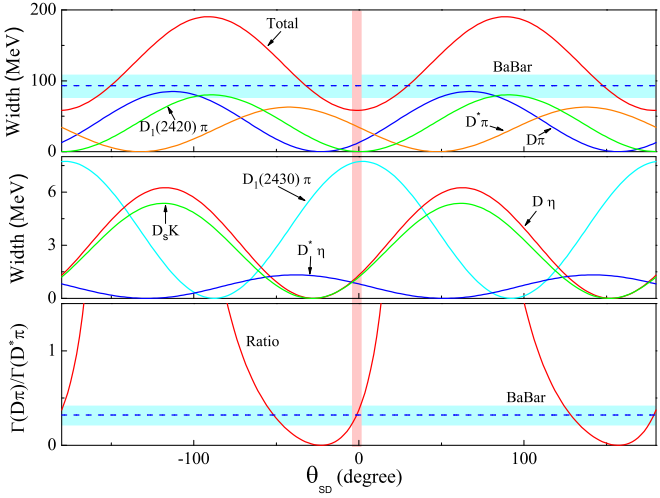}
\caption{\label{fig:2s1d}The $\theta_{SD}$ dependence of the total and partial decay widths and the ratio $\Gamma(D\pi)/\Gamma(D^*\pi)$ of $D^*(2600)$. }
\end{figure}

The $\theta_{SD}$ dependence of the total width, partial decay widths, and ratio
$\mathcal{B}(D^*(2600)\rightarrow D^+\pi^-)/\mathcal{B}(D^*(2600)\rightarrow D^{*+}\pi^-)$ of $D^\ast(2600)$ is shown in Fig. \ref{fig:2s1d}, where two tiny  partial widths $\Gamma(D^*(2600)\rightarrow D^*_sK)$ and $\Gamma(D^*(2600)\rightarrow D^*_2(2460)\pi)$ are not listed.
When taking the range $-3.6^\circ<\theta_{SD}<1.8^\circ$, the theoretical ratio is consistent with the BaBar measurement of Eq. (\ref{26000}). The obtained total width is about 60 MeV which is comparable to the experimental data $\Gamma=93\pm6\pm13$ MeV \cite{delAmoSanchez:2010vq}. We also find that
the main decay modes of $D^\ast(2600)$ is $D\pi$ ($9\sim15$ MeV) and $D^*\pi$ ($32\sim38$ MeV), which also explains why
$D^\ast(2600)$ was first reported in these two decay channels \cite{delAmoSanchez:2010vq}. We need to stress that the small mixing angle $\theta_{SD}$ is expected because of the large mass difference between $D^*(2600)$ and its partner $D^*(2760)$, which is consistent with the suggestion in Ref. \cite{Godfrey:1985xj}.

As the $D$-wave charmed meson with $J^P=2^-$,
$D(2750)$/$D_J(2740)$ is probably either $1D(2^-)$ or $1D'(2^-)$ state, which satisfies the following relation
 \begin{equation}
 \left(
  \begin{array}{c}
   |1D(2^-)\rangle\\
   |1D'(2^-)\rangle\\
  \end{array}
\right )=
\left(
  \begin{array}{cc}
    \cos\theta_{1D} & \sin\theta_{1D} \\
   -\sin\theta_{1D} & \cos\theta_{1D}\\
  \end{array}
\right)
\left(
  \begin{array}{c}
    |1^1D_2  \rangle \\
   |1^3D_2 \rangle\\
  \end{array}
\right),\label{1d}
\end{equation}
where $\theta_{1D}$ is the mixing angle and can be fixed as $\theta_{1D}=-50.8^\circ=-\arcsin(\sqrt{3/5})$ in the heavy quark limit \cite{Godfrey:1986wj,Godfrey:2013aaa,Matsuki:2010zy}.

\begin{figure}
\includegraphics[width=0.49\textwidth]{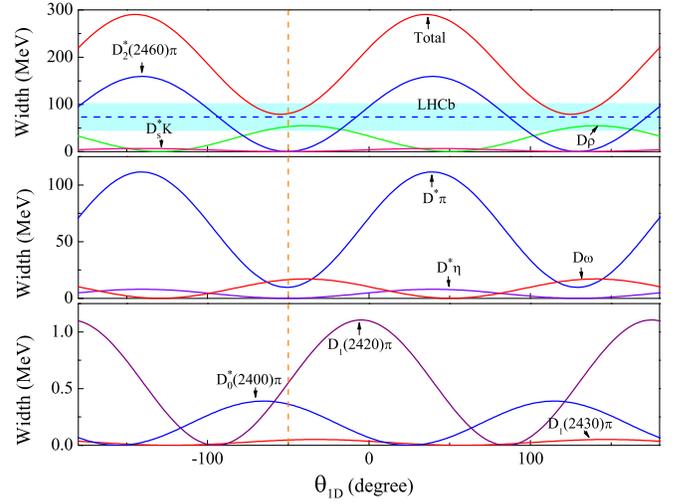}
\caption{\label{fig:1d2}The $\theta_{1D}$ dependence of the total and partial decay widths of $D(2750)$/$D_J(2740)$. Here, the vertical dash line
corresponds to the mixing angle $\theta_{1D}=-50.8^\circ $. }
\end{figure}

The decay modes of $1D(2^-)$/$1D'(2^-)$ are shown in Table \ref{channel2}.
If $D(2750)$/$D_J(2740)$ is $1D^\prime(2^-)$,
the mixing angle dependence of the corresponding partial and total decay widths is given in
Fig. \ref{fig:1d2}.
%Here the calculated total width is consistent with a central value of the experimental data when taking $\theta_{1D}=-50.8^\circ$ predicted by the heavy quark limit \cite{Godfrey:1986wj,Godfrey:2013aaa,Matsuki:2010zy}, which is included in the range $-73.8^\circ<\theta_{1D}<-35.7^\circ$ in Fig. \ref{fig:1d2}.
The range of a mixing angle is obtained as $-73.8^\circ<\theta_{1D}<-35.7^\circ$ in Fig. \ref{fig:1d2} so that the calculated total width is consistent with a central value of the experimental data, which includes the above heavy quark limit, $\theta_{1D}=-50.8^\circ$.
Thus, a $1D^\prime(2^-)$ state is suitable for $D(2750)$/$D_J(2740)$.
In addition, the main decay modes of $D(1D^\prime(2^-))$ are predicted to be $D^*\pi$ ($10\sim25$ MeV), $D\rho$ ($37\sim55$ MeV), $D\omega$ ($12\sim17$ MeV) and $D_2^*(2460)\pi$ ($0\sim25$ MeV).

We need to mention that the widths of $1D^\prime(2^-)$ can be easily transformed into those of $1D(2^-)$, since the width expression for $1D'(2^-)$ with mixing angle $\theta_{1D}$ is equal to that of $1D(2^-)$ with the mixing angle $\theta_{1D}+90^\circ$. Here, we
give more predictions for the missing $D(1D(2^-))$. Its total width can reach $265\sim290$ MeV and its main decay modes are $D^*\pi$ ($96\sim110$ MeV) and $D_2^*(2460)\pi$ ($135\sim160$ MeV) when taking the range $-73.8^\circ<\theta_{1D}<-35.7^\circ$. Here, we take the mass  2737 MeV
for $D(1D(2^-))$ as an input, which is { taken from the experimental data of $D_J(2740)$ \cite{Aaij:2013sza}}.

\renewcommand{\arraystretch}{1.4}
\tabcolsep=11.0pt
 \begin{table} \centering
 \caption{ The calculated partial and total decay widths of $1D$ states with several possible assignments. If a decay channel is forbidden, we mark it by ``--''. As for the channels which depend on other parameters, we label them with ``$\Box$'' and show the widths in Figs. \ref{fig:2s1d} and \ref{fig:1d2}. All values are in units of MeV.}\label{channel2}
\begin{tabular}{ccccccccccccc}
 \\   \toprule[1pt]
Channels        &$1^3D_1$             &$1D(2^-)/1D'(2^-)$ &$1^3D_3$   \\ \midrule[1pt]
$D\pi$          &76.13                & --                &8.47       \\%\bottomrule[1pt]
$D\eta$         &9.01                 & --                &0.31      \\ %\bottomrule[1pt]
$D_sK$          &11.66                & --                &0.17      \\  %\bottomrule[1pt]
$D^*\pi$        &35.16                & $\Box$            &7.05      \\   %\bottomrule[1pt]
$D^*\eta$       &2.68                 & $\Box$            &0.11     \\ %\bottomrule[1pt]
$D_s^*K$        &2.92                 & $\Box$            &0.04     \\%\bottomrule[1pt]
$D\rho$         &26.34                & $\Box$            &0.61      \\ %\bottomrule[1pt]
$D\omega$       &8.87                 & $\Box$            &0.21     \\ %\bottomrule[1pt]
$D^*_0(2400)\pi$&--                   & $\Box$            &--       \\ % \bottomrule[1pt]
$D^*_2(2460)\pi$&0.56                 & $\Box$            &0.63     \\ %\bottomrule[1pt]
$D_1(2420)\pi$  &211.72               & $\Box$            &0.21     \\%\bottomrule[1pt]
$D_1(2430)\pi$  &0.007                & $\Box$            &0.26    \\ \midrule[1pt]
Total width     &385.06               & --                &18.07     \\  \bottomrule[1pt]
\end{tabular}
\end{table}

There are two possible assignments of $D^*(2760)$/$D^*_J(2760)$, i.e. either an orthogonal partner of $D^\ast(2600)$) as given by Eq.~(\ref{sd}) or a $1^3D_3$ state.
Since the mixing angle $\theta_{SD}$ defined in Eq. (\ref{sd}) is quite small, $-3.6^\circ<\theta_{SD}<1.8^\circ$, as shown in Fig. \ref{fig:2s1d}, so it is legitimate to consider that $D^*(2760)$/$D^*_J(2760)$ is dominated by a pure $1^3D_1$ state.

The decay modes of $D(1^3D_1)$ and $D(1^3D_3)$ states are listed in Table \ref{channel2}. %In order to present the numerical results, we take all mixing angles in heavy quark limit.
The theoretical widths of the $D(1^3D_1)$ and $D(1^3D_3)$ states are 385.06 MeV and 18.07 MeV, respectively, both of which deviate from the experimental data, $\Gamma=60.9\pm5.1\pm3.6$ \cite{delAmoSanchez:2010vq} and  $\Gamma=74.4\pm3.4\pm37.0$ MeV \cite{Aaij:2013sza}.
This is the obstacle to assign the $D(1^3D_1)$ and $D(1^3D_3)$ to $D^*(2760)$/$D^*_J(2760)$.

Apart from the problem of their total decay width,
in this work we compare the ratio $\mathcal{B}(D^*(2760)^0\rightarrow D^+\pi^-)/\mathcal{B}(D(2750)^0\rightarrow D^{*+}\pi^-)$ with the BaBar data \cite{delAmoSanchez:2010vq}.
Here, $D(2750)$ is considered as the $1D'(2^-)$ state. If we assign $D^*(2760)$/$D^*_J(2760)$ to the $1^3D_3$ state, the theoretical ratio is $0.34\sim0.86$ with the $-73.8^\circ<\theta_{1D}<-35.7^\circ$ range, which is in good agreement with experimental measurement \cite{delAmoSanchez:2010vq}. As a consequence, $D^*(2760)$/$D^*_J(2760)$ can be tentatively identified as the $1^3D_3$ state. However, the future experimental measurement of the resonance parameters for $D^*(2760)$/$D^*_J(2760)$ is an important topic, which will be helpful for giving a final and definite answer to the assignment of $D^*(2760)$/$D^*_J(2760)$.

Our calculation also shows that the orthogonal partner of $D^\ast(2600)$ should be a broad state with the total width 385 MeV and its main decay channels are $D\pi$, $D^*\pi$, $D\rho$ and $D_1(2420)\pi$. Here, the ratio
\begin{eqnarray}
\frac{\mathcal{B}(D^{*\prime}(1^-)\rightarrow D^*\pi)}{\mathcal{B}(D^{*\prime}(1^-)\rightarrow D\pi)}=0.46
\end{eqnarray}
is also predicted\footnote{
{We also notice recent results of LHCb, where a new resonance $D^*_1(2760)$ \cite{Aaij:2015vea} with spin-1 was observed. Its resonance parameters are
\begin{eqnarray*}
M=2781\pm18\pm11\pm6 \,\mathrm{MeV},\;
\Gamma=177\pm32\pm20\pm7 \,\mathrm{MeV},
\end{eqnarray*}
which are comparable with the theoretical mass (2762 MeV) and width (385 MeV) of $D^{*\prime}(1^-)$. This means that $D^*_1(2760)$ can be the orthogonal partner of $D^\ast(2600)$.
  } }.

\subsection{$3S$ states}

\renewcommand{\arraystretch}{1.4}
 \tabcolsep=6.7pt
 \begin{table*} \centering
 \caption{The calculated partial and total decay widths of $D^*_J(3000)$ and $D_J(3000)$ with several possible assignments. If a decay channel is forbidden, we mark it by "--". All mixing angles are given in the heavy quark limit and all values are in units of MeV. }\label{width1f}
\begin{tabular}{c c c c c c c c c c c c c}
 \\   \toprule[1pt]
Channels        &$3^1S_0$ &$3^3S_1$        &$1^3F_2$          &$1F(3^+)$&$1F'(3^+)$ &$1^3F_4$ &$2^3P_0$  &$2P(1^+)$&$2P'(1^+)$ &$2^3P_2$   \\ \midrule[1pt]
$D\pi$          &--      &13.53             & 26.09            &--                &--       &4.97
&72.51&--&--&1.46\\
$D\eta$         &--      &1.37              & 2.76             &--                &--       &0.29
&6.96&--&--&0.003\\
$D\eta'$        &--      &0.54              & 1.71             &--                &--       &0.02
&1.11&--&--&0.17\\
$D_sK$          &--      &2.01              & 2.78             &--                &--       &0.15
 &7.46&--&--&$1.2\times10^{-4}$\\
$D^*\pi$        &43.17   &25.68             &18.83             &42.23             &8.39     &5.31
&--&61.73&6.46&0.12\\
$D^*\eta$       &2.37    &1.78              &1.76              &3.75              &0.35     &0.21
&--&4.30&1.26&0.23\\
$D^*\eta'$      &0.004   &$1.1\times10^{-5}$&0.06    &$4.9\times10^{-4}$&$8.8\times10^{-6}$ &$7.4\times10^{-5}$   &--&0.90&0.001&0.04\\
$D_s^*K$        &3.10    &2.24              &1.53              &3.06              &0.16     &0.09
 &--&3.77&1.33&0.26\\
$D\rho$         &10.22   &12.16             &19.71             &4.10             &42.86     &2.95
&--&36.44&8.80&13.57\\
$D\omega$       &3.26    &4.31              &6.80              &1.28              &12.99    &0.94
&--&12.56&2.75&4.34\\
$D_sK^*$        &0.004   &0.71              &0.47              &0.01              &0.64     &0.007
&--&1.94&6.75&0.72\\
$D^*\rho$       &2.54    &0.02              &6.41              &16.62             &16.41    &56.94
&138.25&41.94&23.78&5.60\\
$D^*\omega$     &0.83    &0.006             &2.29              &5.08              &5.45     &18.37
 &43.89&13.95&7.78&1.81\\
$D_s^*K^*$      &--      &$1.1\times10^{-4}$&$6.3\times10^{-5}$&--                &--       &$5.8\times10^{-4}$ &3.59&--&--&4.29\\
$D^*_0(2400)\pi$&11.27   &--                & --               &0.23              &0.98     &--
&--&7.08&14.01&-- \\
$D^*_0(2400)\eta$&2.79   &--                & --               &0.002             &0.005    &--
&--&0.39&0.90&-- \\
$D_s(2317)K$     &5.05   &--                & --               &0.004             &0.02     &--
&--&0.84&2.09&-- \\
$D^*_2(2460)\pi$&5.67    &6.93              &11.97             &109.45            &3.27     &2.37
&--&81.44&9.46&5.51\\
$D_1(2420)\pi$  &--      &4.04              &116.51            &1.26              &1.82     &0.33
&11.78&13.03&6.85&7.08\\
$D_1(2420)\eta$ &--      &0.03              &2.21              &0                 &0        &$4.0\times10^{-6}$  &0.32&0.01&0.008&0.02\\
$D_1(2430)\pi$  &--      &1.32              &0.31              &0.12              &0.41     & 1.55
&11.64&9.00&5.03&21.36\\
$D_1(2430)\eta$ &--      &1.40              &$6.8\times10^{-4}$&0                 &0 &$5.1\times10^{-4}$
&0.27&0.006&0.003&0.43\\
$D_s(2460)K$    &--      &2.28              &0.002    &$3.3\times10^{-6}$&$1.2\times10^{-5}$&0.002
&0.57&0.08&0.05&0.88\\ \midrule[1pt]
Total width     &90.28   &80.36             &222.02            &187.20            &93.76    &94.50 &298.35&289.41&97.31&68.89\\\bottomrule[1pt]
\end{tabular}
\end{table*}

The mass spectrum analysis suggests the $D(3^1S_0)$ assignment for $D_J(3000)$ (see the discussion in Sect. \ref{sec3}). Under this assignment, we present the decay behaviors of $D_J(3000)$
in Table \ref{width1f}. Here, its total width is 90.28 MeV which is comparable to the lower limit of the experimental data $\Gamma=188.1\pm44.8$ \cite{Aaij:2013sza}. The partial decay width of
$D(3^1S_0)\to D^*\pi$ is  43.17 MeV which contributes almost 50\% to the total decay width. This is consistent with the fact that $D_J(3000)$ is observed in the channel $D^*\pi$. As a consequence, $D_J(3000)$ can be a good candidate of $D(3^1S_0)$. Besides $D^*\pi$, $D\rho$ and $D^*_0(2400)\pi$ are other two important decay modes.

We further discuss $D^*_J(3000)$ as $D(3^3S_1)$.
In Table \ref{width1f}, the total width of $D(3^1S_1)$ is 80.36 MeV which is pretty close to the lower limit of experimental measurement for $D^*_J(3000)$, $\Gamma=110.5\pm11.5$ MeV \cite{Aaij:2013sza}. The main decay modes of $D(3^1S_0)$ are $D\pi$, $D^*\pi$  and $D\rho$, where the partial decay width of $D(3^3S_1)\to D\pi$ contribute about 17\% to the total decay width which naturally explains why the observed decay mode of $D^*_J(3000)$ is $D\pi$. By the above study, we conclude that $D^*_J(3000)$ as $D(3^3S_1)$ is suitalbe. In addition, we also predict the ratio
\begin{eqnarray}
\frac{\mathcal{B}(D(3^3S_1)\rightarrow D\pi)}{\mathcal{B}(D(3^3S_1)\rightarrow D^* \pi)}=0.53,
\end{eqnarray}
which can be tested in future experiments.

{In this work we do not consider the mixing effect between $2^3S_1$ and $3^3S_1$ since the mass gap between these is as large as 400 MeV as shown in Table \ref{table:spectrum}, which is a different situation from the case of mixing between $2^3S_1$ and $1^3D_1$ discussed in Sect. \ref{m1}.}

\subsection{$1F$ states}

There exist four $1F$ states in the charmed meson family. We list their decay behaviors in Table \ref{width1f}.

As for the $D(1^3F_2)$ state, the mass by the modified GI model is 3053 MeV predicted in Table \ref{table:spectrum}. The $D^*_J(3000)$ recently observed can be a possible candidate of $D(1^3F_2)$ according to the mass spectrum analysis. The main decay channels of $D(1^3F_2)$ are $D\pi$, $D^*\pi$, $D\rho$ and $D_1(2420)\pi$ and the total width is 222.02 MeV, which is about two times larger than the experimental data, $\Gamma=110.5\pm11.5$ MeV \cite{Aaij:2013sza}. Therefore, $D^*_J(3000)$ is not a suitable candidate for $D(1^3F_2)$. Here, the ratio
\begin{eqnarray}
\frac{\mathcal{B}(D(1^3F_2)\rightarrow D\pi)}{\mathcal{B}(D(1^3F_2)\rightarrow D^* \pi)}=1.39.
\end{eqnarray}
is also obtained, which provides a crucial information to further test whether
the $D(1^3F_2)$ assignment to $D^*_J(3000)$ is reasonable.

$D(1F(3^+))$ and  $D(1F'(3^+))$ are mixtures of the $1^1F_3$ and $1^3F_3$ states, i.e.,
\begin{equation}
 \left(
  \begin{array}{c}
   |1F(3^+)\rangle\\
   |1F'(3^+)\rangle\\
  \end{array}
\right )=
\left(
  \begin{array}{cc}
    \cos\theta_{1F} & \sin\theta_{1F} \\
   -\sin\theta_{1F} & \cos\theta_{1F}\\
  \end{array}
\right)
\left(
  \begin{array}{c}
    |1^1F_3  \rangle \\
   |1^3F_3 \rangle\\
  \end{array}
\right),\label{1f}
\end{equation}
where the mixing angle $\theta_{1F}$ can be fixed as $\theta_{1F}=-49.1^\circ=-\arcsin(2/\sqrt{7})$ in the heavy quark limit \cite{Godfrey:1986wj,Matsuki:2010zy} when further discussing their decay properties.

The predicted mass of $D(1F(3^+))$ is 3032 MeV. If we take the assignment $D(1F(3^+))$ to $D_{J}(3000)$, the obtained total decay width is 187.20 MeV, which is in good agreement with the experimental measurement $\Gamma=188.1\pm44.8$ MeV \cite{Aaij:2013sza}. The main decay modes are $D^*\pi$ and $D_2^*(2460)\pi$, which can explain why $D_{J}(3000)$ was first observed by experiment in the $D^*\pi$ channel \cite{Aaij:2013sza}. As a consequence,
$D_{J}(3000)$ can be reasonably regarded as the $D(1F(3^+))$ charmed meson.

As an orthogonal partner of $D(1F(3^+))$, the theoretical mass of $D(1F'(3^+))$ by the modified GI model is 3063 MeV which is about 100 MeV above the experimental data of $D_{J}(3000)$. Thus, we also discuss the $D(1F'(3^+))$ assignment of $D_{J}(3000)$. The results shown in Table \ref{width1f} indicate that $D^*\pi$, $D\rho$, $D\omega$ and $D^*\rho$ are the main decay channels and the total width is 93.76 MeV comparable with the lower limit of the experimental data \cite{Aaij:2013sza}. Therefore, we can not fully exclude the possibility of that $D_{J}(3000)$ is a candidate of $D(1'F(3^+))$.
In order to distinguish possible assignments $D(1F(3^+))$ and $D(1F'(3^+))$ of $D_{J}(3000)$, a precise measurement of the total and partial widths of $D_{J}(3000)$ will be a main task in future experiments.

$D(1^3F_4)$ is a possible assignment to $D^*_J(3000)$ according to only the mass spectrum analysis since the mass of $D(1^3F_4)$ is calculated as 3037 MeV close to the experimental data of $D^*_J(3000)$. In Table \ref{width1f}, we list the decay channels of $D(1^3F_4)$. Here,
$D^*\rho$ and $D^*\omega$ are the main decay channels, and the total width is 94.50 MeV which is consistent with the experimental data $\Gamma=110.5\pm11.5$ MeV \cite{Aaij:2013sza}. However, the decay width of $D(1^3F_4)\to D^*\pi$ is only 4.97 MeV which contributes 5\% to the total decay width. Thus, it is difficult to explain why the observed channel of $D^*_J(3000)$ is $D^*\pi$.
By this analysis, we can conclude that $D^*_J(3000)$ is not a good candidate of $D(1^3F_4)$. We also give extra information of the typical ratio, i.e.,
\begin{eqnarray}
\frac{\mathcal{B}(D(1^3F_4)\rightarrow D\pi)}{\mathcal{B}(D(1^3F_4)\rightarrow D^* \pi)}=0.94,
\end{eqnarray}
although these are not main decay modes.

\subsection{$2P$ states}

In the following, we discuss whether $D_{J}(3000)$ and $D^*_J(3000)$ can be categorized into the $2P$ states.

In Table \ref{width1f}, we list the decay modes of $D(2^3P_0)$, where the main decay modes are $D\pi$, $D^*\rho$ and $D^*\omega$ and the total width is 298.35 MeV which is far larger than the experimental data $\Gamma=110.5\pm11.5$ MeV \cite{Aaij:2013sza}. $D(2^3P_0)$ has mass 2856 MeV predicted in the modified GI model which is about 150 MeV lower than the experimental mass of $D^*_J(3000)$. Hence, if we consider this mass spectrum analysis together with the present calculation of the decay behaviors, the $D(2^3P_0)$ assignment to $D^*_J(3000)$  cannot be supported by our study.

$D_{J}(3000)$ as a candidate of $D(2P(1^+))$ or $D(2P^\prime(1^+))$ is considered here. As mixed states, $D(2P(1^+))$ and $D(2P^\prime(1^+))$ have a relation
\begin{equation}
 \left(
  \begin{array}{c}
   |2P(1^+)\rangle\\
   |2P^\prime(1^+)\rangle\\
  \end{array}
\right )=
\left(
  \begin{array}{cc}
    \cos\theta_{2P} & \sin\theta_{2P} \\
   -\sin\theta_{2P} & \cos\theta_{2P}\\
  \end{array}
\right)
\left(
  \begin{array}{c}
    |2^1P_1  \rangle \\
   |2^3P_1 \rangle\\
  \end{array}
\right)\label{m4}
\end{equation}
with a mixing angle $\theta_{2P}$, where we take $\theta_{2P}=\theta_{1P}=-54.7^\circ$ \cite{Godfrey:1986wj,Matsuki:2010zy} to list the numerical results of the decay widths of
$D(2P(1^+))$ or $D(2P^\prime(1^+))$ in Table \ref{width1f}.

If $D_J(3000)$ is a $D(2P(1^+))$ state, $D^*\pi$. $D\rho$, $D^*\rho$ and $D_2^*(2460)\pi$ are its main decay modes, and its total width can reach
289.41 MeV comparable with the experimental width $\Gamma=188.1\pm44.8$ MeV \cite{Aaij:2013sza}. Although
the predicted mass of $D(2P(1^+))$ is 2853 MeV, which is about 120 MeV lower than experimental value of  $D_{J}(3000)$,
the above results show that there is still a possibility of $D_{J}(3000)$ as $D(2P(1^+))$.

If $D_{J}(3000)$ is $D(2P'(1^+))$, the obtained total width is 97.31 MeV which is comparable to the lower limit of the experimental width of $D_{J}(3000)$. In addition, $D^*\rho$ and $D_0^*(2400)\pi$ are the main decay modes. However, $D^*\pi$ is not a main decay channel since it only contributes 6.6\% to the total decay width, where $D(2P'(1^+))\to D^*\pi$ occurs via a $D$-wave in the heavy quark limit. By remembering that $D_J(3000)$ was first observed in its $D^*\pi$ channel, our results do not favor the $D(2P'(1^+))$ assignment of $D_{J}(3000)$, which is also supported by the mass spectrum analysis since
the theoretical mass of $D(2P'(1^+))$ by the modified GI model is 2885 MeV far below the experimental data.

$D^*_J(3000)$ is not a good candidate of the $D(2^3P_2)$ state, which is concluded by the mass spectrum analysis and the study of its decay behaviors. In the modified GI model, the mass of $D(2^3P_2)$ is predicted to be 2884 MeV, which is inconsistent with the mass of
$D^*_J(3000)$. Under the $D(2^3P_2)$ assignment to $D^*_J(3000)$, the total and partial decay widths are presented in Table \ref{width1f}, where $D\rho$ and $D_1(2430)\pi$ are the main decay modes and the total width is 68.89 MeV which is comparable to the lower limit of experimental measurement $\Gamma=110.5\pm11.5$ MeV \cite{Aaij:2013sza} released by the LHCb Collaboration. However, the process $D(2^3P_2)\to D\pi$ is a subordinate decay channel, which just contributes 2\% to the total decay width. Accordingly, we conclude that $D^*_J(3000)$ is not a good candidate of the $D(2^3P_2)$ state.

In Table \ref{width1f}, we predict abundant information of decay behaviors of the $2P$ states in the charmed meson family, which provides valuable hints to search for the missing $2P$ charmed mesons and to test these meson assignments to $D^*_J(3000)$ and $D_J(3000)$.

\subsection{$2D$ states}

The mass of $D(2^3D_1)$ is 3131 MeV predicted through the modified GI model in Table \ref{table:spectrum}. We show the decay behaviors of $D(2^3D_1)$ in Table \ref{width2d}, which indicates that $D(2^3D_1)$ is a broad state since the obtained total width is 121.75 MeV. Its main decay modes contain $D\pi$, $D^*\pi$ and $D_1(2420)\pi$. We also predict the branching ratio
\begin{eqnarray}
\frac{\mathcal{B}(D(2^3D_1)\rightarrow D^*\pi)}{\mathcal{B}(D(2^3D_1)\rightarrow D \pi)}=0.37.
\end{eqnarray}

The charmed mesons, $D(2D(2^-))$ and  $D(2D'(2^-))$, satisfy the following relation
\begin{equation}
 \left(
  \begin{array}{c}
   |2D(2^-)\rangle\\
   |2D'(2^-)\rangle\\
  \end{array}
\right )=
\left(
  \begin{array}{cc}
    \cos\theta_{2D} & \sin\theta_{2D} \\
   -\sin\theta_{2D} & \cos\theta_{2D}\\
  \end{array}
\right)
\left(
  \begin{array}{c}
    |2^1D_2  \rangle \\
   |2^3D_2 \rangle\\
  \end{array}
\right),\label{2d}
\end{equation}
where  $\theta_{2D}$ is the mixing angle, which in the heavy quark limit we can fix as $\theta_{2D}=-50.8^\circ$  \cite{Godfrey:1986wj,Godfrey:2013aaa,Matsuki:2010zy}.

The $D(2D(2^-))$ has the predicted mass 3115 MeV and has the broad total decay width, whose value can reach 111.67 MeV. Additionally, $D^*\pi$ and $D_{2}^*(2460)\pi$ are the main decay channels, and the decay width of $D(2D(2^-))\to D^*\pi$ can contribute more than 40\% to the total decay width.

With the modified GI model, we get the theoretical mass of $D(2D'(2^-))$ to be 3148 MeV. In Table \ref{width2d}, we list the decay channels of $D(2D'(2^-))$, the total width is 30.17 MeV, and the main decay channels contain $D\rho$ and $D\omega$.

The masses of $D(2D(2^-))$ and $D(2D'(2^-))$ are similar to each other and have the same decay modes but with diferent values. However, we can still distinguish them in experiments in two aspects. Firstly, $D(2D(2^-))$ and $D(2D'(2^-))$ are broad and narrow states. Secondly, in the heavy quark limit $D(2D(2^-))\to D^*\pi$ is a purely $P$-wave decay while  $D(2D'(2^-))\to D^*\pi$ is a purely $F$-wave decay \cite{Godfrey:2013aaa}, which is the reason why the total decay widths of $D(2D(2^-))$ and $D(2D'(2^-))$ are largely different.

The mass prediction of $D(2^3D_3)$ is 3129 MeV in the modified GI model. Since the calculated total width of $D(2^3D_3)$ is 29.33 MeV,
$D(2^3D_1)$ is a narrow charmed meson, where
$D\pi$, $D^*_2(2460)\pi$, $D_1(2420)\pi$ and $D_1(2430)\pi$ are the main decay channels. We also obtain the ratio
\begin{eqnarray}
\frac{\mathcal{B}(D(2^3D_3)\rightarrow D^*\pi)}{\mathcal{B}(D(2^3D_3)\rightarrow D \pi)}=0.29,
\end{eqnarray}
which can be tested in future experiments.

\renewcommand{\arraystretch}{1.4}
\tabcolsep=9.0pt
 \begin{table}[htbp] \centering
 \caption{Decay behaviors of four $2D$ charmed mesons. Values are in units of MeV. }\label{width2d}
\begin{tabular}{cccccccccccc}
 \\   \toprule[1pt]
Channels         &$2^3D_1$     &$2D(2^-)$           &$2D'(2^-)$        &$2^3D_3$\\ \midrule[1pt]
$D\pi$           &36.10             &--             &--                &3.09   \\
$D\eta$          &3.49              &--             &--                &0.11      \\
$D\eta'$         &2.13              &--             &--                &0.003    \\
$D_sK$           &3.46              &--             &--                &0.06    \\
$D^*\pi$         &13.26             &47.39          &0.29              &0.89     \\
$D^*\eta$        &0.99              &3.86           &0.02              &0.001      \\
$D^*\eta'$       &0.02              &0.28           &0.13              &0.03   \\
$D_s^*K$         &0.74              &3.51           &0.04              &$1.3\times10^{-5}$          \\
$D\rho$          &2.19              &5.36           &14.75             &1.02         \\
$D\omega$        &0.71              &1.73           &4.76              &0.32       \\
$D_sK^*$         &0.15              &0.28           &0.01              &0.11  \\
$D^*\rho$        &0.01              &0.29           &1.35              &10.11                \\
$D^*\omega$      &0.004             &0.09           &0.42              &3.35               \\
$D_s^*K^*$       &0.41              &0.47           &0.50              &0.39          \\
$D^*_0(2400)\pi$ &--                &2.54           &2.89              &--     \\
$D^*_0(2400)\eta$&--                &0.18           &0.31              &--     \\
$D^*_{s0}(2317)K$&--                &0.24           &0.36              &--     \\
$D^*_2(2460)\pi$ &7.19              &39.92          &0.60              &2.38              \\
$D^*_{s2}(2573)K$&0.07              &0.37           &0.02              &0.02               \\
$D_1(2420)\pi$   &43.32             &1.41           &0.81              &2.24          \\
$D_1(2420)\eta$  &0.63              &0.10           &0.12              &0.03   \\
$D_1(2430)\pi$   &6.05              &3.36           &2.28              &4.65               \\
$D_1(2430)\eta$  &0.27              &0.12           &0.14              &0.25  \\
$D_{s1}(2460)K$  &0.34              &0.14           &0.16              &0.27 \\
$D_{s1}(2536)K$  &0.22              &0.03           &0.05              &0.007  \\ \midrule[1pt]
Total width      &121.75            &111.67         &30.17             &29.33  \\  \bottomrule[1pt]
\end{tabular}
\end{table}

\section{Summary}\label{sec5}

\renewcommand{\arraystretch}{1.4}
\tabcolsep=2.5pt
\begin{table*}[htbp]
\caption{   {A summary of the possible assignments for $D^*_J(3000)$ and $D_J(3000)$. Comparing the experimental and theoretical data, we rate the possibility of each assignment with stars.}  \label{d3000}}
\centering
\begin{tabular}{lccccc}
\toprule[1pt]%
&Assignments& Mass &Width  &Main channels &    \\ \midrule[1pt]
\multirow{5}{*}{$D^*_J(3000)$ }&$D(3^3S_1)$ & 3015 MeV &80.36 MeV&$D\pi,D^*\pi,D\rho $& $\star \star \star \star \star$ \\
                  &$D(1^3F_2)$ & 3053 MeV &222.02 MeV&$D\pi,D^*\pi,D\rho, D_1(2420)\pi $& $\star $ \\
                  &$D(1^3F_4)$ & 3037 MeV &94.50 MeV&$D^*\rho, D^*\omega $& $\star \star $ \\
                  &$D(2^3P_0)$ & 2856 MeV &298.35 MeV&$D\pi, D^*\rho , D^*\omega$& $\star $ \\
                  &$D(2^3P_2)$ & 2884 MeV &68.89 MeV&$D\rho, D_1(2430)\pi $& $\star $ \\  \midrule[1pt]
\multirow{5}{*}{$D_J(3000)$ }&$D(3^1S_0)$ & 2976 MeV &90.28 MeV&$D^*\pi,D\rho,D^*_0(2400)\pi $& $\star \star \star \star $ \\
                  &$D(1F(3^+))$ & 3032 MeV &187.20 MeV&$D^*\pi, D_2^*(2460)\pi $& $\star \star \star \star $ \\
                  &$D(1F'(3^+))$ & 3063 MeV &93.76 MeV&$D^*\pi, D\rho, D^*\rho, D\omega     $& $\star \star $ \\
                  &$D(2P(1^+))$ & 2853 MeV &289.41 MeV&$D^*\pi, D\rho, D^*\rho,  D_2^*(2460)\pi $& $\star \star $ \\
                  &$D(2P'(1^+))$ & 2885 MeV &97.31 MeV&$D^*\rho, D_0^*(2400)\pi $& $\star $ \\
\bottomrule[1pt]%
\end{tabular}
\end{table*}

Looking at the the observed charmed mesons shown in Table \ref{table:review}, the charmed meson family has become more and more abundant, which has stimulated us to perform a more systematic phenomenological analysis of higher radial and orbital excitations in the charmed meson family with our great interest.

In the present work we have done two major tasks. Firstly, a mass spectrum analysis has been given by adopting the modified GI model, where the screening effect is taken into account. Secondly, the OZI-allowed two-body strong decays of charmed mesons under discussion have been obtained via the QPC model.

In this work, we have revealed the underlying structures of the observed charmed states $D(2550)$, $D^*(2600)$, $D(2750)$, $D^*(2760)$, $D_J(2740)$, $D^*_J(2760)$, $D_J(3000)$, and $D^*_J(3000)$.
{In Table \ref{d3000} we give a summary of possible assignments for $D^*_J(3000)$ and $D_J(3000)$. Comparing the theoretical and experimental data, we rate the possibility of each assignment with stars.}
Additionally, we have provided more abundant properties of these particles including some typical decay ratios and partial decay widths, which are critical to test these possible assignments of charmed mesons.

In the following years, exploration of higher radial and orbital excitations in the charmed meson family will be one of the main projects in Belle, LHCb, and forthcoming BelleII. In this work, we have also predicted some missing charmed mesons, where their masses and decay behaviors have been provided. This information is helpful for experimental study of the missing states in the charmed meson family.
We also expect more experimental observation of charmed mesons in future.

\section*{Acknowledgments}

This project is supported by the National Natural Science
Foundation of China under Grant No. 11222547, No. 11175073, and No. 11375240, the Ministry of Education of China
(SRFDP under Grant No.
2012021111000), and the Fok Ying Tung Education Foundation
(No. 131006).

\end{document}